%% file: sample-sigconf.tex
\documentclass[sigconf, nonacm]{acmart}
\AtBeginDocument{%
  }

\usepackage{booktabs}  % For professional-looking tables
\usepackage{array}  % For custom column widths
\usepackage{multirow}  % For multirow cells
\usepackage{tikz}
\usepackage{smartdiagram}
\usesmartdiagramlibrary{additions}
\usepackage{tikz,tikz-3dplot}
\usepackage{caption}
\usepackage{float}
\restylefloat{table}
\usepackage{placeins}
\usepackage{graphicx}
\usepackage{pgfplots}
\pgfplotsset{compat=1.18}
\usepackage{filecontents}
\usepackage{soul}

\begin{document}

\title[Integrating Cultural Sensitivity in an LLM Chatbot for Reproductive Health]{``\textit{Kya} family planning after marriage \textit{hoti hai}?'': Integrating Cultural Sensitivity in an LLM Chatbot for Reproductive Health
}


\author{Roshini Deva}
\affiliation{%
 \institution{Emory University}
 \city{Atlanta}
 \state{Georgia}
 \country{United States}}
 \email{deva.roshini@emory.edu}

\author{Dhruv Ramani}
\affiliation{%
  \institution{Georgetown University}
  \city{Washington}
  \state{District of Columbia}
  \country{United States}}
  \email{dr973@georgetown.edu}
  
  \author{Tanvi Divate}
\affiliation{%
  \institution{Myna Mahila Foundation}
  \city{Mumbai}
  \state{Maharastra}
  \country{India}}
  \email{tanvi@mynafoundation.com}
  
  \author{Suhani Jalota}
\affiliation{%
  \institution{Stanford University}
  \city{Stanford}
  \state{California}
  \country{United States}}
  \email{suhani@stanford.edu}

\author{Azra Ismail}
\affiliation{%
  \institution{Emory University}
  \city{Atlanta}
  \state{Georgia}
  \country{United States}}
\email{azra.ismail@emory.edu}

\renewcommand{\shortauthors}{Deva et al.}

\begin{abstract}
 Access to sexual and reproductive health information remains a challenge in many communities globally, due to cultural taboos and limited availability of healthcare providers. Public health organizations are increasingly turning to Large Language Models (LLMs) to improve access to timely and personalized information. However, recent HCI scholarship indicates that significant challenges remain in incorporating context awareness and mitigating bias in LLMs. In this paper, we study the development of a culturally-appropriate LLM-based chatbot for reproductive health with underserved women in urban India. Through user interactions, focus groups, and interviews with multiple stakeholders, we examine the chatbot’s response to sensitive and highly contextual queries on reproductive health. Our findings reveal strengths and limitations of the system in capturing local context, and complexities around what constitutes ``culture''. Finally, we discuss how local context might be better integrated, and present a framework to inform the design of culturally-sensitive chatbots for community health.

\end{abstract}





%\newcommand{\AI}[1]{\textcolor{magenta}{[#1 - Azra]}}
%\newcommand{\RD}[1]{\textcolor{red}{#1}}
%\newcommand{\DR}[1]{\st{#1}}
\newcommand{\DR}[1]{}
\newcommand{\RD}[1]{\textcolor{black}{#1}}

 \newcommand{\chiadd}[1]{\textcolor{black}{#1}}
 \newcommand{\chirm}[1]{}

\maketitle
\renewcommand{\thefootnote}{} 
\footnotetext{This is a preprint of the paper accepted at CHI 2025. The final version will be available in the ACM Digital Library.}
\renewcommand{\thefootnote}{\arabic{footnote}}
\input{introduction.tex}
\input{litreview.tex}
\input{methods.tex}
\input{design.tex}
\input{findings.tex}

\input{discussion.tex}

\section{CONCLUSION}
% Our study shows that while the chatbot provides useful health information, it struggles with integrating local dialects, cultural nuances, and behavioral patterns specific to the user's context. This tells the need for more culturally sensitive tools that can effectively bridge the gap between global health technology and local realities. The study emphasizes the importance of incorporating cultural context into the design and evaluation of healthcare chatbots to ensure they meet the needs of diverse populations.
% Future development should be focused on enhancing the chatbot’s ability to recognize local linguistic and cultural variations. Also, feedback from users, healthcare professionals, and developers is essential to refine the chatbot’s responses and improve its cultural sensitivity. By addressing these challenges, we can develop more effective and inclusive public health tools that provide accurate, relevant, and respectful information to all users, eventually improving health outcomes in varied cultural contexts.

This study examined the development of an LLM-based chatbot for providing sexual and reproductive health information to underserved women in urban India. By analyzing user interactions, focus groups, and interviews with multiple stakeholders, we identified both the strengths and limitations of the system in addressing highly contextual and sensitive health queries. Our research highlighted the complexities of delivering culturally sensitive information, and the need for a nuanced understanding of local beliefs and practices. While the chatbot showed promise in delivering relevant health information, challenges remain in ensuring cultural appropriateness and adapting to diverse and evolving community needs. Our paper thus contributes to HCI literature by presenting a framework for designing culturally sensitive LLM-based chatbots. We thus aim to inform future efforts to implement LLMs and chatbots in healthcare settings, particularly with underserved populations.

\section{ACKNOWLEDGEMENTS}
\RD{We thank our participants for generously sharing their time and experiences, which were invaluable to this study. We are also grateful to the external reviewers for their insightful feedback that helped strengthen our work. We want to extend our gratitude to Manvi S and Sumon Kanti Dey from CARE Lab at Emory University, Meet Shah and Harsh Vitra from the technical team, Zeel Mehta from the program team, and Dr. Shraddha Kale Kapile, Associate Director of the Health Vertical at Myna Mahila Foundation for their support throughout the study. This research was made possible with the support of the Google Award for Inclusion Research, and The Agency Fund through Myna Mahila Foundation.}

\bibliographystyle{plain}
\bibliography{sample-base}
%\bibliographystyle{ACM-Reference-Format}

\input{appendix.tex}

\end{document}

%% file: introduction.tex
\section{INTRODUCTION}

There has been significant interest in the potential for LLMs in healthcare---to support health information seeking, data summarization, delivery of care services, and more ~\cite{singhal2023large, shahsavar2023role, zhou2023survey}. Several of these efforts focus on leveraging this technology to improve healthcare delivery, including in underserved settings in the Global South ~\cite{kimblack2022, tsengunder, 10.1007/978-3-030-60114-0_10}.
Most LLMs, however, are primarily trained on data and literature in English from the Global North, and inherit racial, gender, and cultural biases persistent in these databases ~\cite{8418155, denecke2021evaluation}. 
They can fail to recognize local dialects, cultural nuances, and speaking patterns, especially for populations that are less represented online and do not speak English, furthering the technology access gap ~\cite{adilazuarda2024measuringmodelingculturellms, ondrejova2024can, li2024culturellm}. 
Several ongoing efforts have focused on both adapting existing models and developing entirely new language models (including more efficient and sustainable ones) to improve performance in non-English languages ~\cite{10.1007/978-3-030-60114-0_10,  Ożegalska-Łukasik_Łukasik_2023,10.1145/3639856.3639890}. Our paper aims to contribute to this landscape by informing the design of \textit{culturally sensitive} LLMs in healthcare, by considering the design of a chatbot in an underserved setting in India.

% This includes several ongoing efforts on the use of LLMs reproductive health \cite{}. 
% and chatbots are designed to cater to the medical systems in the west, they render useless when evaluated in different contexts. 

Our interest in the role of culture \chiadd{in LLM-driven text generation} is guided by its close relationship to language and \chiadd{health} communication; how it both shapes and is shaped by it.
% \textbf{}
\chiadd{LLMs frequently struggle to accurately interpret health-related queries for users from diverse backgrounds, partly due to insufficient training data ~\cite{banerjee2024navigating}. This can lead to misunderstanding culturally specific health practices, which we hope to address through our work. } % Additionally, LLMs face persistent challenges with question awareness, resulting in inappropriate responses to medical queries \cite{yang2024llms}.
The term ``culture'' has a long history of being studied in relation to technology use within the field of Human-Computer Interaction (HCI), though it has rarely been defined concretely (e.g. ~\cite{ge2024culture,clemmensen2010overview,lofstrom2010culture}). 
% It has been used as a catch-all phrase to describe research done with non-WEIRD (Western and mostly white, Educated, Industrial, Rich, Democracies) populations, and 
It has frequently been used interchangeably with local context, community, nationality, religion, as well as race and ethnicity---\chiadd{and to distinguish from most research with white, privileged, or university-educated populations}. 
In this paper, \chiadd{we draw on a rich body of work in health communication literature on the role of culture}. We define culture as \textit{``shared values, norms, codes, roles, and assumptions that shape a group's beliefs, attitudes, and behavior through their interactions in and with their environments''} ~\cite{griffith2024cultural}.
\chiadd{Building on this understanding, we rely on Resnicow et al.'s definition of cultural sensitivity as \textit{``the extent to which ethnic, cultural, and other factors are incorporated in the design, delivery, and evaluation of health communication, health promotion materials, and health promotion programs''}~\cite{resnicow1999cultural}. However, LLMs offer unique challenges to design in comparison to more traditional group and individual-level health messaging, which we discuss in this paper.}

We focus on Sexual and Reproductive Health (SRH), a topic that has been highly taboo in contexts around the world \cite{giritharan2020socio,likith2024exploring}, and that surfaces many of the tensions around culture in the development of LLM-based healthcare interventions. 
\RD{SRH has been deeply shaped by cultural norms---reflected in normative pressures, media discourse, and spousal dynamics \cite{feriani2024systematic}. 
%At times restricting open conversations and access to accurate information, especially among women \cite{giritharan2020socio,likith2024exploring}. 
Stigmas surrounding ``cleanliness'' during menstruation and sexual health behaviors can lead to social exclusion and discrimination \cite{espinosa2019breaking}, while harmful practices like virginity testing in some communities may reinforce restrictive societal expectations \cite{robatjazi2015virginity}. The fear of judgment and shame can prevent individuals from seeking medical care for sexually transmitted infections (STIs) or unplanned pregnancies \cite{cook2014reducing}, impacting both physical and mental health.} 
%Moreover, reproductive and contraceptive choices are influenced by normative pressures, media discourse, and spousal dynamics \cite{feriani2024systematic}.}
\chiadd{Finally, colloquial language may be used locally to refer to SRH concerns instead of medical terms in English.} \RD{LLMs risk magnifying these barriers if they fail to account for the culturally specific language people use to navigate stigma.}

We partnered with a Non-Governmental Organization (NGO) called Myna Mahila Foundation (Myna) based in Mumbai (India). We studied the design and use of an LLM-based chatbot that Myna had developed for providing information and services on SRH. \chiadd{Myna used a multilingual LLM (preview model of GPT-4) that offers support for Hinglish (transliterated Hindi, \textit{i.e.} Hindi written with the Roman script, with code-mixing between Hindi and English). Based on initial testing with community members, Hinglish was identified as the most preferred language for chat interactions and was used in the final chatbot.}
The chatbot was extensively tested with a largely Hindi-speaking, but culturally and religiously diverse group of women \chirm{workers} in an underserved migrant urban community in Mumbai, within Myna's outreach area.
We \chiadd{primarily} studied 2118 message logs (question-answer pairs) of the \chirm{workers'} \chiadd{women's} interactions with the chatbot, analyzing both the types of questions asked and the chatbot's effectiveness in responding to them. We also supplemented this with data from two focus groups with the\chirm{workers} \chiadd{women}, and focus groups, interviews, and online WhatsApp and Slack discussions with healthcare professionals, and technology developers at Myna to understand design decisions that were made to improve the cultural sensitivity of the chatbot. 
% We analyzed this data to consider the strengths and limitations of the chatbot in delivering culturally appropriate health advice.
We paid special attention to the role of stigma and taboos, linguistic context, nuances around social norms, and the role of family dynamics. 
% Our research thus aims to offer a perspective on the design of LLM-based chatbots that meet the needs of a community.

Our paper is structured as follows. We begin by presenting literature on LLMs and chatbots in healthcare and on culturally sensitive design. This is followed by details about our study context and methods, and the design of the chatbot. 
Our findings reveal the strengths and limitations of the system in capturing local context and complexities around
what constitutes ``culture''.
We also identify the elements of culture that emerged as being relevant and highly critical to address in an LLM-based intervention. 
We then discuss the implications of these findings for the future development of personalized LLM-driven tools in public health in diverse settings.
Finally, we consider how we might integrate the local context into chatbots, and present a framework to inform the design of culturally-sensitive LLMs for community health.

%% file: litreview.tex
\section{RELATED WORK}
Below we present literature on LLMs and chatbots in healthcare, design principles for chatbots, and research on cultural sensitivity in HCI and with LLMs. 

\subsection{Information-Seeking and Chatbots in Healthcare}
\chiadd{HCI literature has well-established that information seeking and processing are deeply influenced by people's daily lives and their social contexts--both in terms of how one finds and searches for accurate or verifiable information, and how they process and apply it \cite{savolainen1995everyday,savolainen2022assessing}. 
In high-stakes environments such as healthcare, getting access to timely and accurate information becomes even more critical. HCI has long documented how people make sense of health information \cite{raj2019clinical,young2019girl,nakikj2023alerts}, such as through social media \cite{mcdowall2024sensemaking}. 
% X \cite{}. 
For underserved communities specifically, information gaps are amplified by systemic barriers like limited digital literacy \cite{arias2023digital, ehrari2022digital}, distrust in automated systems \cite{ehrari2022digital}, contradictory nature of certain cultural and healthcare practices \cite{azongo2014complementary, craffert1997opposing},  and general difficulties in navigating through existing healthcare systems \cite{pervez2023systemic, ahmed2001barriers}. Prior work by Yadav et al. \RD{ and Wang et al.} has aimed to bridge this gap using chatbots designed to meet the needs of breastfeeding women and adolescent sex education while keeping in mind these constraints \cite{yadav_feedpal_2019, wang2022artificial}.}  
\RD{By incorporating local beliefs and considering social dynamics like family influence \cite{yadav_feedpal_2019, wang2022artificial}, they aim to provide culturally congruent health education.
%Their chatbot Feedpal, designed to serve as the first point of contact for breastfeeding mothers \cite{yadav_feedpal_2019}, and SnehAI,  designed for SRH awareness among youth \cite{wang2022artificial}, adapted their responses to align with local beliefs. By incorporating culturally contextualized responses and considering social dynamics such as family influence their approach highlights how chatbots can be adapted for high-impact, community-driven health education. 
%Yadav et al. emphasize the significance of culturally aware chatbot design in healthcare. 
}
 
% These patterns become more nuanced as one seeks information about more sensitive subjects such as health. 
% These studies further strengthen our proposal of incorporating sociocultural contexts into LLM-based chatbots to account for diverse information-seeking patterns across different communities. 
% Incorporation of such contextualized healthcare information-seeking patterns \cite{savolainen1995everyday, savolainen2022assessing} onto LLMs thus becomes very important in providing care and help. 

Recent studies in HCI and health informatics show a growing interest in utilizing chatbots in configurable healthcare interventions to improve patient engagement due to their anthropomorphic communication styles~\cite{singhal2023large, shahsavar2023role, zhou2023survey, sallam2023chatgpt, parmar2022health}. Usage of LLM-based chatbots could possibly mark a new era of digital patient engagement and care delivery ~\cite{ruggiano2021chatbots, shahsavar2023role}. The use of LLM-based chatbots for healthcare started with a lot of attention on aiding conversations around mental health ~\cite{song2024typing, althoff2016large, yang2023towards,10.1145/3613904.3642761}, virtual caregiving ~\cite{ruggiano2021chatbots, miura2022assisting, wang2021evaluation} and personalized health information delivery ~\cite{harrington2023trust, almalki2020health, skjuve2018chatbots}. LLMs have been particularly beneficial for these applications, as they can provide desired information while allowing for open-ended questions and conversations ~\cite{jovanovic2020chatbots, holmes2019usability, dolianiti2020chatbots,10.1145/3544548.3581503}. Their multilingual and personalization capabilities ~\cite{kocaballi2019personalization, liu2022effects, ait2023power,10.1145/3613905.3651093} also allow tailoring of interactions to the individual linguistic background and their health literacy levels ~\cite{marzo2024bridging}. Studies have shown that these models, when paired with factual clinical knowledge, have the potential to offer contextually relevant information in response to user's queries, making them suitable for complex healthcare interactions. 

The rapid advancement of LLMs in healthcare requires frameworks to address ethical concerns ~\cite{song2024typing, sepahpour2020ethical, chen2023chatbots, 10.1145/3613904.3642385}.
Major issues include ensuring clinically-accurate medical responses ~\cite{abbasian2024foundation, goodman2023accuracy}, privacy concerns ~\cite{li2023security, kanter2023health} and potential biases ~\cite{kim2023assessing}. In response, the World Health Organization has released ethics and governance guidelines for the safe use of LLMs in healthcare~\cite{whoguidelines}. Though these chatbots might be able to help people who have no other options at their disposal ~\cite{song2024typing}, evidence suggests that they have notable risks~\cite{song2024typing, sepahpour2020ethical}. Hence, there has been a growing research interest around the design ~\cite{wilson2022development, tsengunder, griffin2023chatbot} and evaluation of chatbots in healthcare ~\cite{cascella2023evaluating, denecke2021evaluation}, and documentation of ethical challenges that they may pose ~\cite{sepahpour2020ethical, li2023security}. 

\subsection{Design of Chatbots}
The design of healthcare chatbots prioritizes usage of high-quality clinical knowledge ~\cite{singhal2023large, denecke2021evaluation}, security, safety ~\cite{li2023security, abbasian2024foundation} and accurate comprehension of the user's query. The HCI community has placed a lot of emphasis on the importance of appropriate user experience ~\cite{hornbaek2017technology, hassenzahl2006user}, by assessment and incorporation of emotional intelligence ~\cite{bilquise2022emotionally, ghandeharioun2019towards, pamungkas2019emotionally} and adaptability ~\cite{nadarzynski2023but} to these chatbots. These factors are crucial to the design of the bot given the risks involved in healthcare interactions ~\cite{sepahpour2020ethical, li2023security} and the empathy required ~\cite{alam2022examining, seitz2024artificial, seitz2021empathic}. As with other HCI research, there is a strong emphasis on user-centered design ~\cite{abras2004user, soloway1994learner}. One of the primary challenges is ensuring clinical accuracy while maintaining conversational fluency based on the user’s linguistic context ~\cite{chavesling, spilnertalk2021} and health information literacy ~\cite{abreu2024utilization}. This not only involves design modifications during development but also tailoring the medical knowledge base ~\cite{8418155, chung2019chatbot}, incorporating real-world scenarios and patient-doctor dialogues ~\cite{chaix2019chatbots, chung2019chatbot} to capture the nuances of medical communication. 

The user experience is also designed to encourage user’s trust, with incorporation of empathy ~\cite{seitz2021empathic, seitz2024artificial,10.1145/3613904.3642336}, accessibility ~\cite{habicht2024closing, stanley2022chatbot} and clarity.  A persona is provided to the bot based on the use case and the user’s context, and personalization of responses is promoted~\cite{nissen2022effects, hwang2021applying}. Studies have shown that users are more likely to engage with and trust bots that present a professional appearance while maintaining a warm, empathetic tone ~\cite{seitz2021empathic, seitz2024artificial}. 
\RD{Yadav et al. and Wang et al. emphasized the significance of culturally aware chatbot design in fostering comfort, trust, and cultural relevance ~\cite{yadav_feedpal_2019, wang2022artificial}. SnehAI featured a female persona based on a popular television character and adopted a narrative approach aligning with the Indian context \cite{wang2022artificial}, whereas FeedPal modeled its female persona after trusted maternal health workers, to answer user questions from the knowledge base \cite{yadav_feedpal_2019}.}

%Acknowledging that women may be reluctant to discuss maternal and sexual health due to stigma, the significance of culturally aware chatbot design in healthcare was emphasized \cite{wang2022artificial, yadav_feedpal_2019}. To foster comfort, trust, and cultural relevance, ensuring women felt at ease engaging in discussions around sensitive topics---a female persona modeled after trusted maternal health workers, Accredited Social Health Activists (ASHAs), and a persona based on popular television character that takes a narrative approach aligning with the indian context were featured \cite{wang2022artificial, yadav_feedpal_2019}.}

LLM-based chatbots are pivotal for the level of personalization they provide, which has historically not been possible ~\cite{kocaballi2019personalization}. They leverage the user's profile, interaction history and their health information to provide medical information that is the most appropriate for that user~\cite{openai2024gpt4technicalreport}. Personalization also helps in tailoring the interactions based on the user's conversation style ~\cite{kocaballi2019personalization}, preferences, and their health literacy level, hence adapting to their tone and approach. Explainability is also promoted and utmost importance is given to addressing the risks and ethical concerns~\cite{ai4030034, shuchaoreview}. Building on top of personalization and explainability, there is work on incorporating the user’s local and cultural context ~\cite{adilazuarda2024measuringmodelingculturellms, 10.1145/1878450.1878481, Ożegalska-Łukasik_Łukasik_2023}. This becomes even more important when working with underserved communities ~\cite{harrington2023trust, kimblack2022, tsengunder}. 

\subsection{Cultural Sensitivity}
\chiadd{In health communication literature, Resnicow et al. define cultural sensitivity as \textit{``the extent to which ethnic, cultural, and other factors are incorporated in the design, delivery, and evaluation of health communication, health promotion materials, and health promotion programs''} \cite{resnicow1999cultural}. They go on to highlight two primary dimensions of cultural sensitivity---surface structure and deep structure \cite{resnicow1999cultural} \textit{Surface structure} focuses on the inclusion of visible and identifiable elements of a specific population, such as language, imagery, music, food, clothing, and other cultural symbols that resonate with the group.
In contrast, \textit{deep structure} delves into the underlying cultural factors that shape behaviors, including values, beliefs, norms, and stressors. For example, it involves tailoring content to reflect core cultural values like family commitment, spirituality, or respect for elders, as well as addressing unique stressors such as discrimination or racism. By incorporating these deeper cultural drivers, health messages and interventions can more effectively influence behaviors and outcomes \cite{resnicow1999cultural}. While these theoretical understandings of cultural sensitivity pre-date the use of LLMs (which raise a unique set of challenges), they offer us a starting point to expand and build a framework for integrating cultural sensitivity in LLM-based interventions in healthcare settings.}

%\RD{Yadav et al. and Wang et al.’s work closely align with these principles. They accounted for surface structure by ensuring that their chatbots, supported code-mix and voice input, enhancing accessibility for users with diverse literacy skills. More significantly, their approach addressed deep structure by considering social influences on breastfeeding practices, particularly the role of mothers-in-law in shaping maternal health decisions, and tackled social stigma surrounding SRH discussions. Instead of opposing traditional beliefs, both employed affirmative interaction; however, their implementation differed. Feedpal used structured guidance and gentle cues, while SnehAI relied on narrative-driven engagement to counter misinformation with medically validated responses by identifying cultural myths. Despite their strengths, these approaches have restricted adaptability, pointing to the need for further research into chatbot models that can flexibly and contextually respond to cultural sensitivities.}

The capability of LLMs to appropriately respond to the cultural context of the user is a crucial aspect of personalization needed for healthcare information seeking ~\cite{adilazuarda2024measuringmodelingculturellms, 10.1007/978-3-030-60114-0_10, Ożegalska-Łukasik_Łukasik_2023}. In particular, communities that are underserved and less represented online may have less trust in AI and the healthcare system ~\cite{harrington2023trust, kimblack2022, tsengunder}. Taking into account linguistic diversity and responding to the user’s query in their preferred language is also critical. But being sensitive to the culture goes beyond the linguistic translation ~\cite{orosoo2024enhancing, heim2013}. It requires a deep understanding of cultural differences, their local dialects, nuances, behavioral patterns, and their sociocultural context ~\cite{10.1007/978-3-030-60114-0_10}. Despite advancements in current LLMs, they often struggle with cultural misunderstandings given that the training data is predominantly in English from Western contexts ~\cite{tsengunder, 10.1145/1878450.1878481, griffith2024cultural}. Recent work by Adilazuarda et \textit{al.} has defined proxies of culture, which can be mapped to certain measurable aspects of LLM’s responses and language ~\cite{adilazuarda2024measuringmodelingculturellms}. Based on modeling and linguistic analysis, they give guidelines and techniques to improve cultural and linguistic sensitivity ~\cite{griffith2024cultural, 10.1007/978-3-030-60114-0_10, article}. 

% \chiadd{While there are various challenges and misalignments in usage of LLMs in these contexts, there are some specific to healthcare information access \cite{yang2023large} which require additional focus. There are differences in how health conditions are understood and described across cultures \cite{hogg2010cultural, bakic2012conceptual}, hence designing these system requires varying the preferences of array of various qualitative parameters \cite{bakic2012conceptual, bakic2018communicating} like directness \cite{singleton2009understanding, bakic2018communicating}, expectations \cite{singleton2009understanding, bakic2018communicating} , level of information \cite{singleton2009understanding, bakic2018communicating}, the tone of interactions \cite{singleton2009understanding, bakic2018communicating}, etc.}

Within the HCI community, researchers have defined cultural sensitivity as the extent to which ethnic, cultural, and local contexts are incorporated in the design, delivery, and evaluation of these chatbots ~\cite{adilazuarda2024measuringmodelingculturellms, shi2024culturebank, li2024culture}. This may entail aligning with different communication styles, e.g. directness, tone, dialect ~\cite{adilazuarda2024measuringmodelingculturellms, ondrejova2024can, li2024culturellm}, and adapting to what is preferred locally and culturally. It also includes recognizing cultural variations in family roles in healthcare, including gender norms ~\cite{anderson2003culturally, vaughn2009cultural,cerezo2023healthcare}, differing levels of healthcare access, and health and data literacies~\cite{almalki2020health, 10.1007/978-3-030-60114-0_10}. 
% Recognizing cultural barriers to healthcare becomes crucial, especially if the goal is to improve access. 
% Based on all of these considerations, user research is done with direct user involvement throughout the process ~\cite{abras2004user, soloway1994learner}. 
Based on this understanding, appropriate interface and interaction design can be carried out, with localization and accessibility considerations ~\cite{seitz2021empathic, habicht2024closing}. 
These cultural contexts are kept at the forefront during evaluation ~\cite{10.1007/978-3-030-60114-0_10, adilazuarda2024measuringmodelingculturellms}. 

%% file: methods.tex
\section{METHODS}
Our study aimed to understand the cultural appropriateness of an LLM-based chatbot designed to provide information on sexual and reproductive health (SRH). 
% to women in urban India, and identify elements that constitute ``culture'' in the local context. 
We analyzed log data from \chiadd{women from an underserved community in Mumbai who were recruited by} \chirm{workers at}Myna to test the application. We also conducted focus groups with them\chirm{workers}, and interviews and focus groups with developers, healthcare professionals, and the program team at Myna. Below, we detail the \chiadd{background of the women from the community,} participant recruitment process, data collection, and analysis methods. We obtained approval to conduct the research from the Institutional Review Board at Emory University in the United States.

\begin{figure*}[h]
  \centering
\includegraphics[width=1\textwidth]{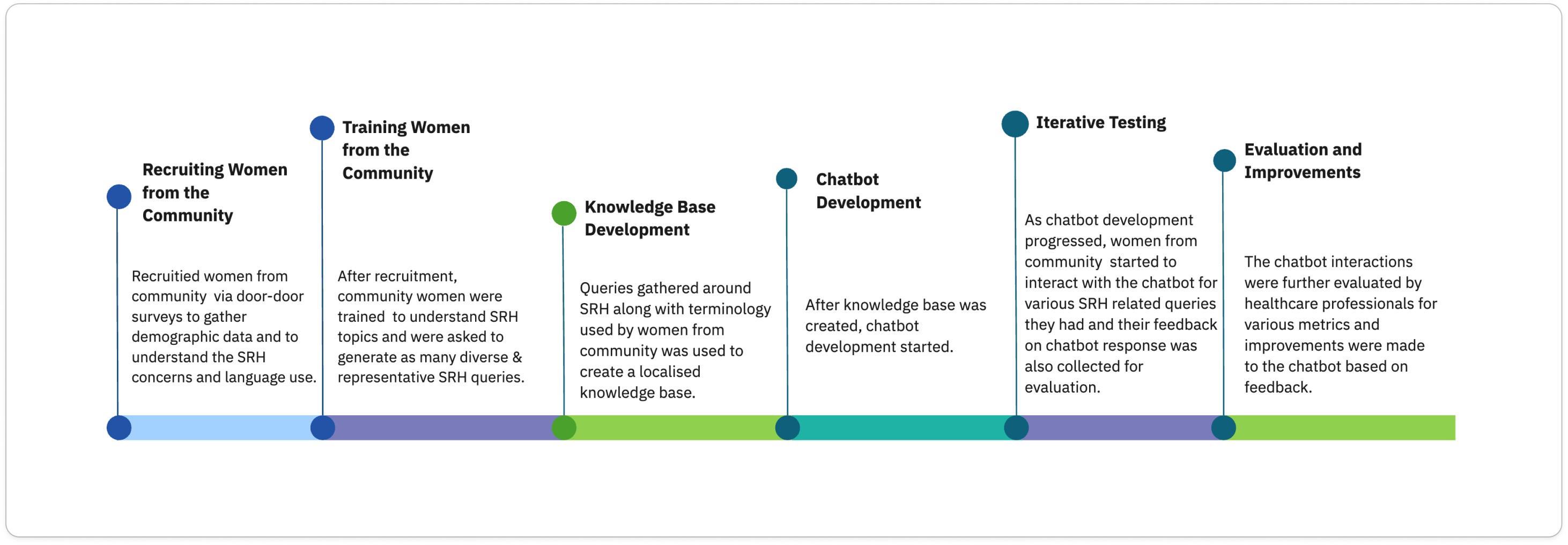} 
  \caption{\textbf{Various phases of the study}.}
  \label{fig:methods}
  \Description{Six different phases of our research study. Starting with Recruiting Women from the Community Via door-door surveys, Training Women from the community to understand SRH topics, Knowledge Base Development using queries gathered from women, Chatbot Development followed by Iterative testing of the chatbot with community women, and finally Evaluation and Improvements.}
\end{figure*}

\subsection{Study Context and Community Engagement}
We were introduced to women from the community, technology developers, and healthcare professionals through Myna, a non-profit working on sexual and reproductive health in Mumbai, India, since 2016.
\chiadd{The women who participated in testing the chatbot in our study were initially employed by Myna as part of a micro-tasking employment program called Rani Jobs. Despite several of them having previously engaged with Myna through their menstrual health programs, they were initially recruited via door-to-door surveys by Myna's program team to ensure a diverse group of participants representing a broad spectrum of SRH concerns and language use. Many of them had little to no prior knowledge about SRH topics and chatbots when recruited. As a representative sample, the women were well-positioned to understand and articulate SRH issues women face in their community. }

\chiadd{Prioritizing a community-centric approach, Myna engaged the women across various phases of the chatbot development process, including the pre-development phase, iterative testing, and evaluation. 
Figure \ref{fig:methods} outlines all the phases and the community's involvement.
The women were recruited even before the chatbot was developed and were paid to create a question bank to help potentially fine-tune the LLM (an approach that was later abandoned in favor of pre-trained models with RAG, detailed in Section 4).  
They were called to Myna's local centers for digital training to learn to use a spreadsheet to record their questions, and to build awareness on SRH topics such as family planning, pregnancy, infertility, and menstrual health management to help with question generation.
Follow-up sessions were set up at Myna's center after the chatbot was developed to onboard them to the platform.}
% Peer-to-peer learning and interactive sessions with health experts was used to create a safe and supportive environment for discussing such sensitive topics. 
% , to provide more context into who the women were, their specific roles in each phase, and how their active participation was key in shaping the study’s outcomes. However, the chatbot interactions presented in this paper focus on the data collected during the testing and evaluation phase. Informed consent was obtained from all participants, and confidentiality was strictly maintained to protect their identities and personal information. The women were compensated for their contribution, and treated by Myna as paid contract workers.}
%The study adopted a deliberate, iterative approach, with input from women at every phase playing an important role in shaping the dvelopemnt process. %we will not be including details about the pre-development and development phase of the knowledge base and chatbot.

\subsection{Data Collection}
\chirm{The recruitment strategy aimed to include a diverse group of women to reflect a range of SRH concerns and language use. The community women were recruited for this study via door-to-door surveys by the Myna team. Being members of the same community, they were capable of understanding the issues faced by women in these areas.}  
% \subsection{Data Collection}
\chiadd{As part of the testing phase, women from the community were first invited to digital centers where they engaged with the system under the supervision of the program team. The team monitored these interactions and assisted with prompt generation if needed. Observations made by the team during this phase were recorded for analysis.
% Participants showed a strong interest in asking questions and seeking information from the chatbot. They asked their family members and neighbors about their concerns and started prompting those questions in the interactions from their homes. 
The testing phase is iterative and ongoing, however, for the scope of our study, we specifically analyzed interactions with the chatbot over a period of three months, from late December 2023 to mid-March 2024.} 
All interactions during this period were logged in a database, including timestamps, questions asked, chatbot responses, conversation and message identifiers, and the language used. 
Additionally, feedback from women on various aspects such as response length, understandability, the difficulty of terms used in responses, and latency was collected by the Myna team to support improvements to the knowledge base.
% In the later phase, all the logged responses were reviewed and commented on by healthcare professionals for metrics such as medical accuracy, social context accuracy, interpretation of questions, breaking of misconceptions, and simple medical terminology. 
% Additionally, quantitative feedback from the women on various aspects such as response length, understandability, the difficulty of terms used in responses, and latency was collected by the Myna team to support improvements to the knowledge base.
\chiadd{After the chatbot testing, two focus groups were conducted with the women to understand their backgrounds and perspectives better.}
Our participants included a diverse group of women, in terms of age, religion, family status, and education. Their demographic data is presented in Table \ref{tab:demo}.

We also interviewed two developers, one of the healthcare professionals, and the program team virtually,
and recorded the conversations for further analysis. % on Google Meet 
The healthcare professional we interviewed is the Associate Director of the health vertical of Myna, and has been associated with the organization for six years. 
The healthcare professionals had varying qualifications; all of them had either a diploma certificate in gynecology and obstetrics or more than ten years of experience in the field of sexual and reproductive health while working with underserved communities. 
\chirm{They not only have professional medical training but have had experience working with underserved communities, are able to understand what interventions help those people, and understand the language and cultural taboos that exist among women in such communities.}
They were responsible for testing and evaluating the medical accuracy and contextual relevance of the chatbot's responses, \chiadd{leveraging their experience and understanding of SRH language and taboos}. The developers had experience working on AI applications, they were responsible for enhancing the chatbot in multiple phases. 
All the healthcare professionals were women, while the chatbot developers were men.
% We also conducted interviews and focus groups with healthcare professionals, Myna's program team, and technology developers.  
We also had access to WhatsApp and Slack chat communications with five healthcare professionals and two chatbot developers. 
They actively communicated on these channels throughout the processes of chatbot development, testing, evaluation, and improvement. These chats were also recorded for analysis. 
As we analyzed the chatbot interactions, the other data sources provided important context on complexities designing a culturally sensitive chatbot.

% We analyzed the log data from the interactions between these workers and the chatbot and conducted two focus groups with the workers to better understand their backgrounds and perspectives.
% We also conducted interviews and focus groups with healthcare professionals, Myna's program team, and technology developers.  The WhatsApp and Slack communications among them were also recorded for analysis. 
% As we analyzed the log data, the other sources of data provided important context on the complexities of designing a culturally sensitive chatbot.

\chirm{As part of the testing process, community workers were called to Myna's in-house digital centers to onboard to the chatbot in group sessions, and to encourage them to ask questions about SRH by creating a safe and supporting environment for discussing sensitive topics.
% as well as get training on SRH, considering cultural taboos around the topic. 
Peer-to-peer learning was used to create a safe and supportive environment for discussing sensitive topics. 
Once the community workers felt comfortable asking questions related to SRH, they were encouraged to interact with the chatbot by asking questions on family planning, pregnancy, and menstrual health-related topics using Hinglish (or transliterated Hindi). This interaction helped in generating a training dataset for the chatbot. The community workers played a crucial role throughout the chatbot's development, participating from the early stages of dataset generation to the validation phase. Their involvement ensured the chatbot was trained with relevant, culturally sensitive queries, enhancing its effectiveness in addressing SRH topics. Initially, they were invited to digital centers where they engaged with the system under the supervision of the program team. The team monitored these interactions and assisted with prompt generation. Observations made by the team during this phase were also recorded for analysis. 

In the later phases, participants showed a strong interest in asking questions and seeking information from the chatbot. They asked their family members and neighbors about their concerns and started prompting those questions in the interactions from their home. Interaction with the chatbot continued over a period of nearly three months, from late December 2023 to mid-March 2024. }
% During this period, all interactions were logged in a database, including timestamps, questions asked, chatbot responses, conversation and message identifiers, and the language used. All the logged responses were later reviewed and commented on by healthcare professionals. 
% They were asked to evaluate for the metrics of medical accuracy, social context and terminology used, 

\subsection{Data Analysis}
The analysis phase was done in several stages. First, the transcribed log data from user chatbot interactions and WhatsApp and Slack communications were translated into English. However, the Hinglish text was analyzed alongside the English text to retain meaning. A total of 2118 question-answer pairs from the message logs were further analyzed using open coding to identify key issues related to the chatbot’s effectiveness and cultural sensitivity.
% \chirm{ We conducted an inductive analysis and coding process on the chatbot log data to identify patterns and common themes in user queries and chatbot responses. The coding process followed an iterative approach, beginning with line-by-line coding to identify the SRH topics of the user prompts. We further filtered out the prompts and responses that had a cultural relevance. }
\chiadd{Given that the log data was in the form of spreadsheets and involved analyzing chatbot conversations in both Hinglish and English, we chose to do manual analysis over using analysis software. This ensured that each user interaction, including prompts and chatbot responses, was examined in detail in time sequence, capturing complexities that automated tools might overlook.} 
%Our analysis aimed to capture not only the explicit content but also the implicit meanings and cultural relevance situated in the interactions, ensuring a comprehensive understanding of the data. 
We conducted an inductive analysis and iterative coding process to identify patterns and common themes in user queries and chatbot responses, and with the interview data.
We started with line-by-line coding, and categorized user prompts by SRH topics and type of questions. 
\chiadd{The coding was conducted by the first two authors independently. Discussions were held with the whole team regularly to compare, refine, and consolidate the codes.
In case of conflicting codes, the last author weighed in and helped reach a resolution through discussion.
Through this iterative process, codes were carefully evaluated in the context of surrounding text, preserving the intent of user prompts and the chatbot's responses. This collaborative approach helped ensure inter-coder reliability.}

\chiadd{In particular, we analyzed user prompts and chatbot responses for the role of culture. To identify relevant cultural components, we relied on the definition of culture presented earlier, as ``\textit{shared values} (e.g. community cohesiveness and support), \textit{norms} (e.g. community dynamics, laws, and regulations), \textit{codes} (e.g. colloquial language on SRH instead of medical terms), \textit{roles} (e.g. gender roles), \textit{and assumptions} (e.g. medical consensus) \textit{that shape a group's beliefs} (e.g. on the importance of vaccination), \textit{attitudes} (e.g. taboos around SRH), \textit{and behavior} (e.g. menstrual practices, diet)'' \cite{griffith2024cultural}.
We considered both the literal meaning of the text and its potential meaning in the broader sociocultural context and analyzed to what extent this was understood by the chatbot.}  
During this process, the WhatsApp and Slack conversations, along with data from interviews and focus groups, provided context on how healthcare professionals perceived the accuracy and cultural sensitivity of chatbot responses.

\subsection{Study Limitations}
A limitation of our study is that we could not link user IDs to their demographic data during analysis.
Also, the participants were women from the community who were employed and monetarily compensated by Myna specifically to generate data, which may have influenced user engagement. This testing stage, which took place both on Myna's premises and in the \chirm{workers'} \chiadd{women's} homes, is critical before implementing with the community as the chatbot's performance was uncertain. In future research, we plan to test the chatbot with community members accessing the bot only from their homes, to better understand organic uptake and address user concerns before launching the application more broadly.

\subsection{Positionality}
Our team includes individuals from academia and the program team at Myna. All of us are of Indian origin, working or living in India or the United States. As a group, we come from diverse cultural and religious backgrounds. One of us identifies as a cis-man, and the rest as cis-women. We all have lived experiences and observations around sexual and reproductive health in Indian settings, with a collective experience of almost two decades working on gendered health and wellbeing, with and without use of technology. \chiadd{
% Our lived experiences and professional insights have shaped our understanding of the sociocultural and systemic factors affecting reproductive health outcomes. 
Our lived experiences with the cultural and social context and professional insights guided chatbot development and interpretation of findings.
% , as this type of analysis necessitates an understanding of the cultural and social context. 
We acknowledge that our own positionality may have introduced certain assumptions and biases. 
In particular, we are committed to centering women's agency when making decisions about their own reproductive health. 
To address biases, we actively engaged healthcare professionals and women from the community to check our assumptions during analysis.}

\begin{table}
\centering
\small
\begin{tabular}{l l l l l}
\toprule
\textbf{Age} & \textbf{Religion} & \textbf{Married Status} & \textbf{Children} & \textbf{Education}  \\ 
\toprule
32 & Hindu & Married & 2 & 10th  \\ 
32 & Hindu & Married & 2 & SY.BA \\ 
- & Hindu & Married & 0 & 10th  \\ 
19 & Hindu & Married & 0 & SY.BA  \\ 
33 & Hindu & Married & 2 & 12th  \\ 
27 & Muslim & Married & 1 & 10th  \\ 
28 & Muslim & Married & 2 & 11th \\ 
37 & Buddhist & Married & 1 & FY.BA \\ 
34 & Muslim & Married & 2 & FY.BA \\ 
45 & Hindu & Married & 2 & 12th \\ 
38 & Muslim & Married & 4 & 10th \\ 
32 & Hindu & Married & 2 & 12th \\ 
37 & Muslim & Married & 2 & 10th \\ 
30 & Muslim & Married & 2 & 10th  \\ 
31 & Hindu & Married & 2 & 12th  \\ 
34 & Hindu & Married & 2 & Graduation  \\ 
42 & Hindu & Married & 2 & 12th \\ 
28 & Hindu & Married & 2 & BA Completed  \\ 
\bottomrule
  \end{tabular}\\
    % \vspace{10pt}
 \caption{\textbf{Demographic information about our study participants.} Fields marked as “-” indicate that the information was not collected. Participant IDs were assigned to ensure anonymization. FY.BA and SY.BA refers to the completion of one or two years respectively of a three-year BA degree.}
  \label{tab:demo}
  \end{table}

%% file: design.tex
\section{DESIGN OF THE CHATBOT}

\begin{figure}[h]
  \centering
  \includegraphics[width=0.25\textwidth]{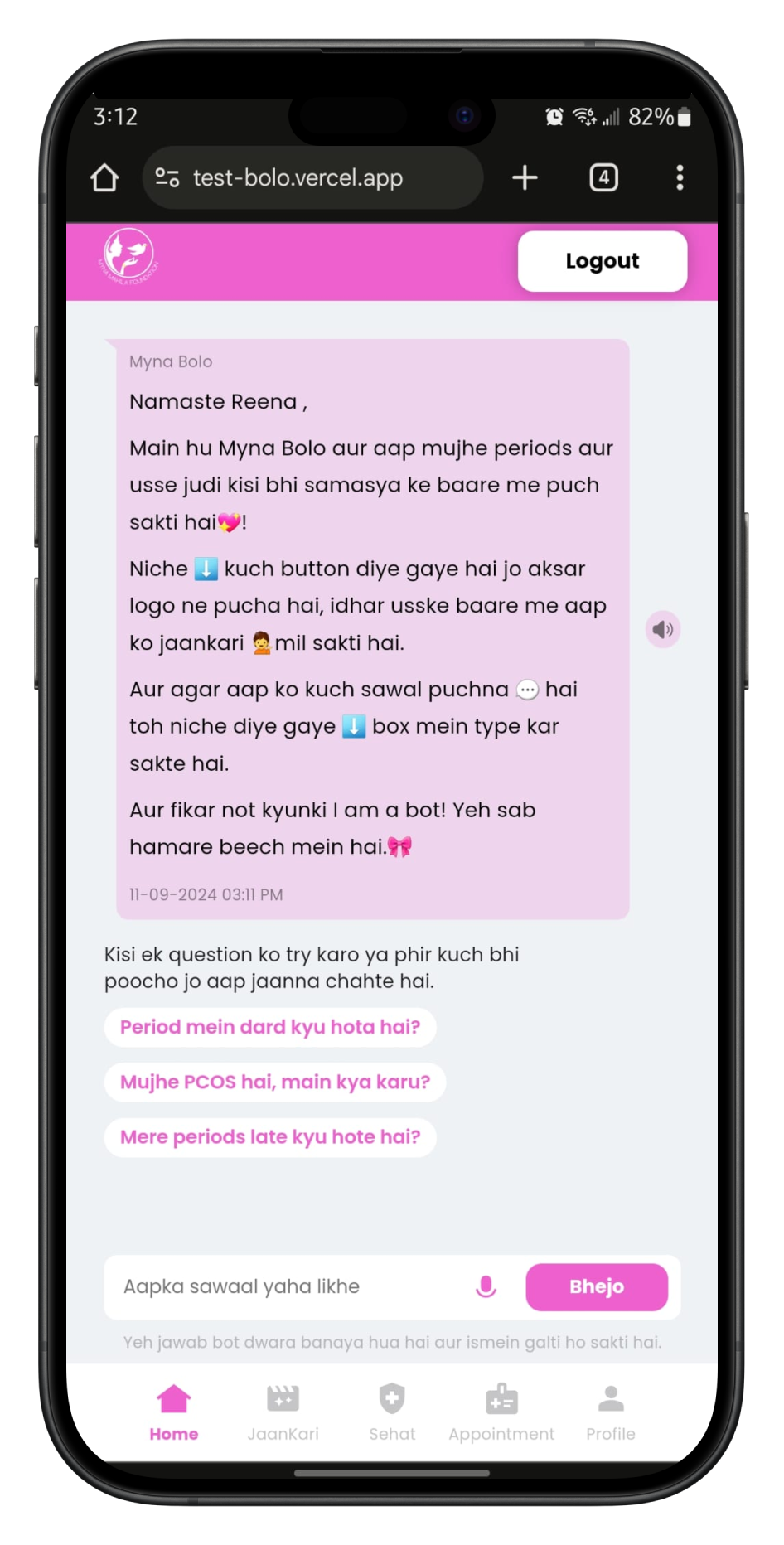} 
  \caption{\textbf{Chatbot Interface on Mobile Web Platform}. The figure presents a greeting in Hinglish and an explanation of what the chatbot is capable of. It also presents buttons below the text with suggested questions to click on to get started. On the right is a button with the speaker icon which plays automated text-to-speech on clicking.}
  \label{fig:interface}
  \Description{Chatbot Interface screenshot on Mobile Web Platform. The top right end has a logout button for the chatbot. The interface has a welcome message in Hinglish for the user, followed by some question suggestions. Read aloud button next to the welcome note. The interface consists of an input field to type in a prompt for the chatbot. Microphone icon in the input field for the user to speak. The menu bar is at the bottom end of the screen.}
\end{figure}

\begin{figure*}[h]
  \centering
  \includegraphics[width=1\textwidth]{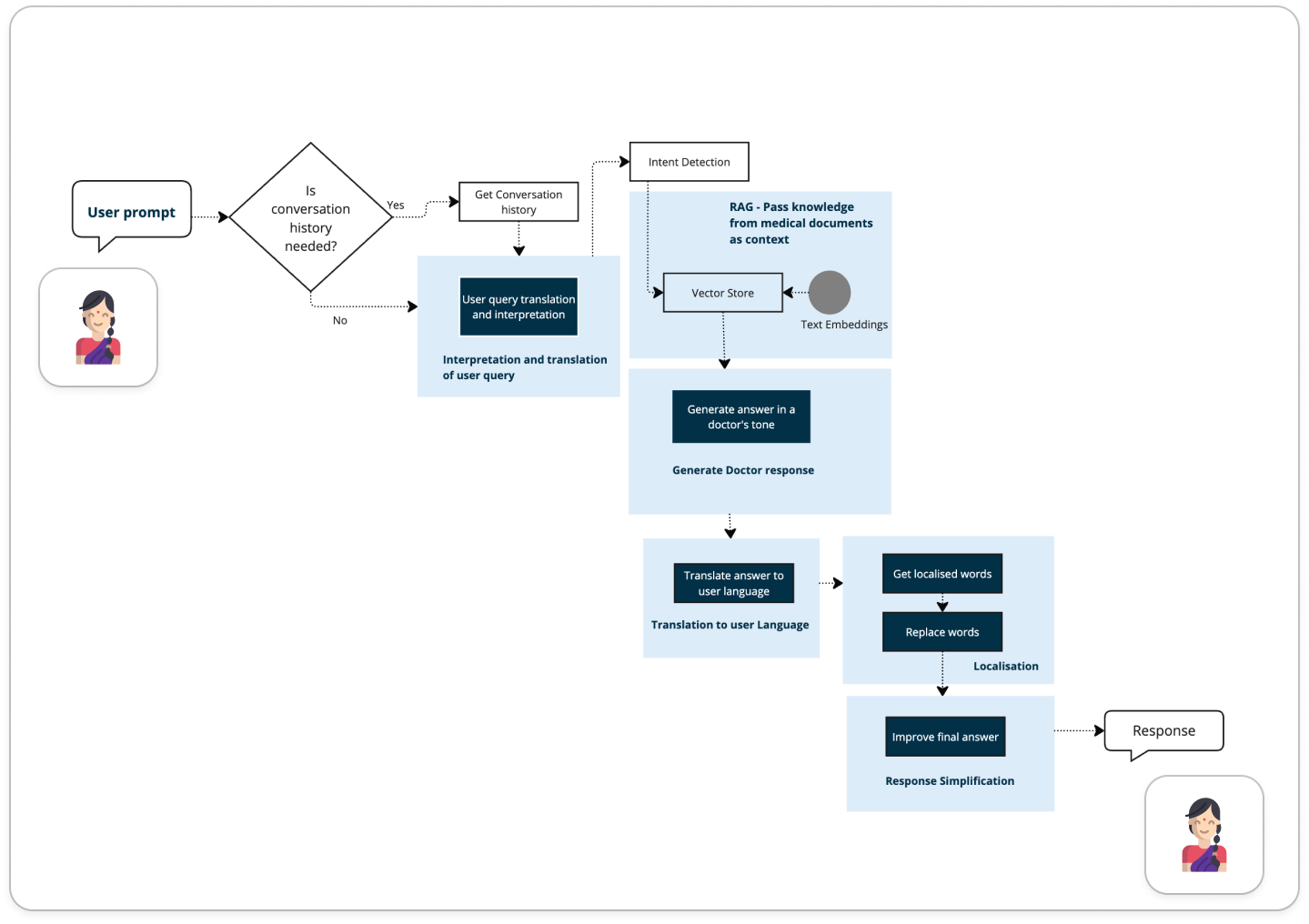} 
  \caption{\textbf{Final system architecture of the chatbot}. The three stages of the chatbot flow include---Translation module, Generating the medical answer, and Localization module. The translation module involves OpenAI’s LLM model which interprets and translates the user’s query in Hinglish into English. Generating the medical answer involves generating the medical answer from the chatbot’s knowledge base by prompting the LLM with a predefined prompt and translating it back to the user language. The localization module involves replacing complex medical words with colloquial terms.
}
  \label{fig:design1}
   \Description{Three stages of the chatbot: Translation module, Generating the medical answer, Localization module. Translation module involves OpenAI’s LLM which interprets and translates the user’s query in Hinglish into English. Generating the medical answer involves generating the medical answer from the chatbot’s knowledge base by prompting the LLM with a predefined prompt and translating it back to the user language. Localization module involves localization process of medical answers generated where words are replaced with their localized counterparts.}
\end{figure*}

We now share more details about the final design and architecture of the chatbot. The chatbot's design was aimed towards providing non-judgmental, confidential, and medically accurate advice on sexual and reproductive health (SRH). It underwent a series of iterative improvements to enhance its effectiveness, cultural sensitivity, and overall user experience in four phases with different GPT models, including a fine-tuned GPT-3.5 model, GPT-3.5, and GPT-4. The final system was developed using the preview model of GPT-4. The chatbot was hosted on a custom web application, as shown in Figure \ref{fig:interface}. It interfaced in Hinglish \chirm{(or transliterated Hindi)} which was common in digital interactions in the study context. In addition to receiving responses to questions as text, users could choose the text-to-speech button to listen to an audio recording of the text. This was included to accommodate users with a range of literacies. The final architecture of the bot consists of three main stages. A brief overview of the system's final architecture is described in Figure \ref{fig:design1}.

In the \textit{first stage}, the user's query in Hinglish, is interpreted and translated into English using OpenAI's LLM translation services. This process leverages LLMs with prompts designed to accommodate the linguistic and cultural backgrounds of the target population. The translated query is then passed further to capture the correct medical context around the query. This is done by consulting medical documents (contextualized and developed) by MMF through a retriever model like RAG (Retrieval Augmented Generation) that draws on a vector store of text embeddings of these documents. \chiadd{Our approach follows a widely adopted methodology, which involves the following steps: First, the documents in english are split into smaller chunks and stored as vector embeddings in a vector database. A query is then used to find similar chunks in the database, and the retrieved chunks are added to the prompt.} This ensures the responses are medically informed and tailored to the user's needs and context, and reduces the possibility of hallucinations and inaccuracies in the text generated.

The \textit{second stage} involves generating the medical answer from the bot's knowledge base. This is facilitated by prompting an LLM with relevant background information and instructions. The prompt asks the LLM to generate a response as a seasoned female gynecologist and obstetrician from India with over 25 years of expertise (see Table \ref{tab:prompt} in Appendix for the detailed prompt). Emphasis is placed on the importance of cultural sensitivity, empathy, and compassion when interacting with underserved Indian women who may lack in-depth biological knowledge. The LLM is instructed to engage in a step-by-step conversation, mirroring the approach that a doctor would take. This includes identifying potential causes for any mentioned ailments, systematically gathering patient information, and providing precise guidance based on a clear understanding of the user's problem. The LLM's response is then passed to the third stage for localization and final output generation.

In the \textit{third stage}, the medical answer from the LLM undergoes a localization process to ensure its cultural and linguistic relevance. The system adapts the language of the response to match local dialects and usage patterns, employing a service that performs fuzzy matching to find and replace words with their localized counterparts. Before delivering the final response to the user, the chatbot simplifies the text to ensure grammatical correctness and ease of understanding. 
The length of the chatbot responses ranged from 25 words for the shortest to 393 words for the longest. On average, responses contained 123 words. Throughout the workflow, various safeguards are implemented to ensure the chatbot operates within its intended scope. For instance, the chatbot is instructed to avoid recommending tests outright, refrain from prescribing medications, and emphasize the importance of consulting a healthcare professional for further diagnosis and treatment. 

%% file: findings.tex
\section{FINDINGS}
% In this section we discuss the findings of our study, focusing on cultural elements of the local context. 
We now present learnings from our analysis of both the questions asked by the users, and the chatbot responses to sensitive and highly contextual queries on sexual and reproductive health.
\DR{We interpret these with the support of data gathered and analyzed from interviews, focus groups, and online conversations with the healthcare professionals, developers, and the program team at Myna.} 
\RD{We organize our findings by first noting broader trends around information seeking behaviors and focusing on \textit{surface structure} ~\cite{resnicow1999cultural}---such as the topics, timings, and types of questions asked, and the language of interaction---based on user log data. We then moved our focus to the \textit{deep structure}---the underlying cultural factors that shape behaviors, including values, beliefs, norms, and
stressors.
This led us to note the role of cultural beliefs and stigma, and legal and social dynamics in shaping interactions. 
% Here, we drew on the definition of culture stated earlier, \textit{i.e.} \textit{``shared values, norms, codes, roles, and assumptions that shape a group’s beliefs, attitudes, and behavior through their interactions''} \cite{griffith2024cultural}. 
Based on the chat timestamps, we further reviewed Slack and WhatsApp conversations within the development and program teams and used insights from interviews and focus groups. This gave us more context on when and why the system failed to respond to a user query correctly, and to understand the role of design decisions.} 
%We structured our findings using the definition of culture we adopted. We grouped our findings based on the analysis of chatbot interactions, to identify indicators of cultural relevance in user engagement. These themes were then contextualized with insights from interviews, focus groups, and online conversations with the healthcare professionals, developers, and the program team at Myna. By integrating these diverse data sources, we aimed to evaluate how cultural factors and design changes over time influenced chatbot performance and to interpret the patterns observed in user interactions.}
Our findings highlight the strengths and limitations of the system in appropriately recognizing and responding to diverse cultural elements in its interactions. 
% We identified several culturally sensitive elements from the user interactions presented graphically in Fig.1.

% In this section, we present sample user messages and bot responses. 
In several places, we have included both the original Hinglish text as well as the translated English text. This is done in an effort to retain the cultural context and showcase how questions and responses were framed, which can be lost in translation. For the English translations, we have aimed to provide literal translations as much as possible, which may result in some oddly framed sentences, but try to retain the original style and meaning of the user messages. Finally, when quoting from WhatsApp or Slack conversations between team members at Myna, we have shared the original text (which was mostly in English), including any typos. All names mentioned in the paper are pseudonyms.

\subsection{Understanding Information-Seeking Behaviors Around SRH}
A major concern for Myna was ensuring that the chatbot took into account the literacies and socioeconomic realities of the participants. 
Below we first summarize patterns we observed in the questions that users were interested in asking of the chatbot. We then reflect on the conversation styles of both users and chatbots and the challenges those could present. We then discuss whether the chatbot could provide easily comprehensible text to users. We also consider how demographic factors such as income and education played a role in conversations.

\subsubsection{Topics, types, and timing of questions asked by users}
We analyzed the log data to identify the various topics and types of questions asked by users. These categorizations are presented in Tables \ref{tab:topic} and \ref{tab:type}, with the number of queries that fell under each category. Myna had instructed the \chiadd{women recruited from the community} \chirm{workers} to generate questions about family planning, pregnancy, and menstrual health---which are typically the focus of women's health programs in India. However, our analysis revealed that the questions asked on the chatbot extended beyond these topics, indicating a strong interest and need in other SRH areas. Table \ref{tab:topic} shows that questions also revolved around the use of contraceptive pregnancy, reproductive anatomy, and marriage and relationships. \chiadd{
% While all the topics can be classified under one broad category, they have been distinguished to account for the specificity and frequency of user queries, allowing for a more detailed and clear analysis. 
Table \ref{tab:type} presents the types of questions asked by users, and most were basic conceptual inquiries (577), followed by 504 number of more advice/opinion queries.
This differentiation was made to understand the SRH literacy level of the women and serve as a precursor to evaluate the cultural appropriateness. For instance, \textit{``Condom Kya hota hai?''} translated as \textit{`What is condom?'} has been categorized as a basic informational question where as \textit{``15 saal ki ladki kaise karegi family planning?''} translated as \textit{``How will a 15 year old girl do the family planning?''} was categorized as a question on norms and ethics. Examples of how user queries were categorized by topic and the type of question are presented in Table \ref{tab:examples} in the Appendix.}
% However, due to space constraints, we have not included examples for every possible combination of topic and question type.
Most users asked knowledge-seeking questions (over half of all user queries), suggesting a strong desire for basic educational content on the topic. However, our analysis of the types of questions asked indicated that a significant number of questions focused on concerns about their specific situation with their partner or in-laws, safety concerns around contraceptives, social norms around sex and family planning, misconceptions, and more (Table \ref{tab:type}). 

Our analysis of users' prompts showed a variability in length. While some users provided very short queries, others provided more elaborate ones. The length of user prompts ranged from a minimum of 3 words to a maximum of 70 words, with an average length of 11 words. We found that the chatbot could handle user prompts of varying lengths.
We also analyzed the timestamps from the user interactions log data to determine the times of day when users were most active. We identified that interactions began as early as 2 AM. Interaction activity peaked between 6 AM and noon, with nearly 1,900 of 2118 queries being asked during this time frame. After noon the interaction activity gradually declined for the rest of the day. This may be because of other responsibilities that they were expected to manage at home, with them having most personal quiet time before their family members wake up and after they leave home for work and school.

% We observed the need for providing information that enhances their understanding of the topics they are inquiring about. 

% 

\begin{figure}[h]
  \centering
  \includegraphics[width=0.5\textwidth]{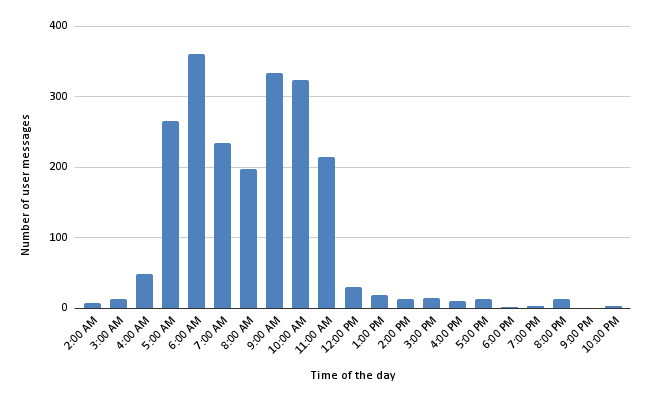} 
  \caption{\textbf{Timing of questions asked by users.} The graph summarizes the number of queries received at different times of the day. The highest activity was between 5AM and 12PM.}
  \label{fig:timeslot}
 \Description{The bar graph plots the number of user messages against the time of day, with the X-axis representing the time in hourly intervals from 2:00 AM to 10:00 PM and the Y-axis representing the number of user messages ranging from 0 to 400. The highest message activity is observed between 6:00 AM and 12:00 PM, peaking at around 7:00 AM and 12:00 PM, with over 300 messages during these hours. There is minimal activity after 1:00 PM, with a significant drop in user messages throughout the afternoon and evening. Very few messages are sent between 2:00 AM and 5:00 AM, as well as after 7:00 PM.}
\end{figure}

\begin{table*}[h]
\centering
\small
  \begin{tabular}{>{\raggedright\arraybackslash}p{3cm} >{\raggedright\arraybackslash}p{9cm} >{\raggedleft\arraybackslash}p{2.5cm}}
\toprule
\textbf{Topic} &  \textbf{Description} & \textbf{Number of queries} \\
\toprule
Contraceptive methods & Information about \chirm{various ways to prevent pregnancy} \chiadd{various non-permanent contraceptive methods} & 657 \\
Family planning & Guidance on planning pregnancies &541 \\
Sexual Health &Information on safe sexual practices\chiadd{, including protection from HIV and STDs} & 258 \\
Pregnancy & Information on conception, and health during pregnancy&227 \\
Sterilization &  Permanent methods of contraception & 182 \\
Reproductive anatomy & Structure and function of the male and female reproductive systems. & 99 \\
Menstrual health & Concerns about menstrual cycles&80 \\
Abortion &  Information on the medical termination of pregnancy & 34 \\
\chiadd{Fertility support} \chirm{IVF} & Information about assisted reproductive technology,  \chiadd{such as IVF} & 17 \\
Miscarriage & Information on natural loss of pregnancy & 04 \\
Follow-up & Generic followup questions &09 \\
\bottomrule
\textbf{Total} &  &\textbf{2118} \\
  \end{tabular}
    % \vspace{10pt}
 \caption{\textbf{Categorization of SRH Topics Covered by Users.} Categorization of topics with the description of how they were grouped. \chiadd{The Follow-up category includes prompts asked by users in direct response to a previous interaction or by clicking on a follow-up question suggested by the chatbot. Examples for categorization are presented in the Table \ref{tab:examples}.} }
  \label{tab:topic}
  \Description{Categorization of sexual and reproductive health topics with the description of how they were grouped and the number of queries asked for each topic. Contraceptive methods is the most asked topic with 657 questions and Miscarriage was the lowest with 4 questions.}
  \end{table*}
\begin{table*}[h]
\centering
\small
  \begin{tabular}{>{\raggedright\arraybackslash}p{3.5cm} >{\raggedright\arraybackslash}p{8.5cm} >{\raggedleft\arraybackslash}p{2.5cm}}
\toprule
\textbf{\chirm{Topic} \chiadd{Type of Questions}} &\textbf{Description}& \textbf{Number of queries} \\
\toprule
\chiadd{Basic Conceptual Inquiries} & \chiadd{Inquiries focused on basic information about SRH topics, such as definitions of SRH terms} & \chiadd{577} \\     
\chiadd{Complex Queries} & \chiadd{Inquiries that go beyond basic SRH knowledge, such as use of a specific contraceptive} & \chiadd{438} \\  
Advice/Opinion & Seeking personal recommendations, \chiadd{and guidance about making informed choices in personal or health-related issues} &\chiadd{504}\chirm{220} \\
Health and Safety concerns & Safety issues and health risks \chiadd{, including concerns around hygiene and safe sexual practices such as preventing HIV and STDs} & \chiadd{365}\\ 
Misconception &  Clarifying misunderstandings or beliefs &137 \\
Norms and Ethics & Concerns about social norms and ethics, \chiadd{including expectations on gender roles} &\chirm{18}\chiadd{68}\\
Healthcare Access & Challenges in accessing healthcare due to financial or social constraints &20 \\
Follow-up & Generic followup questions & 09 \\
\bottomrule
\textbf{Total} &  &\textbf{2118} \\
  \end{tabular}
    % \vspace{10pt}
 \caption{\textbf{Types of Questions Asked by Users.} Types of questions with description based intent of the user prompt. \chiadd{The Follow-up category includes prompts asked by users in direct response to a previous interaction or response provided by the chatbot. } }
  \label{tab:type}
  \Description{Categorization of types of questions with descriptions based on intent of the user prompt. Knowledge-seeking questions were the most asked questions with 924 and decision-making questions were the least asked with 17. Follow-up questions were generic questions that did not fit into any category. Examples for categorization are presented in the Table \ref{tab:examples}.  }
  \end{table*}

\subsubsection{Communication style of users and the chatbot}
We paid close attention to the grammar, sentence construction, choice of both health and non-medical terms, and dialects used by the users and the chatbot. 
These attributes play an important role in interpreting the user prompt and generating a response that is easily understandable by the user.
We noticed that initially, the chatbot responses were designed to provide information to user queries directly without acknowledging the local social norm. After a system design update, a predefined response template was used to improve the interaction by incorporating cultural expression, as shown in Figure \ref{fig:interface}. The addition of ``Namaste'' gave the interaction a more culturally respectful tone, acknowledging the user before giving the information. 
We also learned that the tone of the chatbot had to be updated. For instance, early in the chatbot development process, Dr. Preeti shared that the chatbot should be using \textit{``aap, aapka, aapko''} [seen as a respectful and formal way of addressing someone in the second-person] instead of \textit{``tu, tujhe''} (informal ways of addressing someone).

In our focus groups with workers, they also expressed frustration when the chatbot's responses were overly generic or dismissive, particularly when it directed them to consult a doctor without providing sufficient information. For instance, users seeking guidance on family planning may expect more than a referral to a healthcare provider. They might want actionable insights or advice that considers societal expectations, financial constraints, and community norms. 
Though suggesting talking to a medical professional seems a reasonable way to ensure no over-reliance on chatbot responses, in practice, it resulted in users feeling dissatisfied with the response. This was because they were using the chatbot as they did not have another channel available to them.
Balancing medical recommendations with culturally sensitive, actionable information is thus key to maintaining user trust and engagement. To address this, Myna integrated the option of doing a teleconsultation, rather than just suggesting that the user talk to a doctor.

We also noted that few follow-up questions were asked by participants, even when encouraged by the chatbot. For instance, we noted that one user asked a question on the possible reasons for not getting pregnant (see Table \ref{tab:reasons} in the Appendix). The chatbot asked follow-up questions to gather specific details about the user’s situation, such as changes in menstrual cycles, weight fluctuations, or any existing health issues like thyroid problems or PCOS. However, the user responded by saying that they did not have any information and requested an explanation of the possible reasons.
Despite the lack of context, the chatbot effectively responded to the query by providing a generic list of potential reasons for not getting pregnant. 
This shows the system's capability to give relevant information even when a user was not willing to provide further details. 
% This variability in the context-seeking approach across interactions shows the chatbot’s limitation.

\subsubsection{Adapting to local language patterns}
We noted that user queries frequently included grammatically incorrect sentence structures (in both Hindi and English), spelling errors, and the use of terms that may be local to the context rather than a medical term.
By design, the chatbot did not account for regional dialects or script variations in the user's prompt. Before the generation of the response with the LLM, another prompt was fed to the LLM to ``correct'' the spelling and grammatical errors in the query.
This approach resulted in the chatbot being able to generate medically accurate answers in most cases, but it could occasionally result in misunderstandings of user queries if context was lost through this approach. For instance, the chatbot struggled to understand words with slight differences in spelling or meaning. Dr. Sheela pointed out one example in a WhatsApp conversation with the Myna team: \textit{``English response to Hindi question.the bot failed to distinguish the words `rokna' and `rukna' (miscarriage and abortion)''} showing that the chatbot's language processing capabilities are not efficient in handling nuanced context-specific meanings. %As a result, chatbot misinterpreted the user query, and provided inaccurate and irrelevant responses. 
%We observed that there are instances where the chatbot used English words in its responses even when the user’s query is entirely in Hindi. We noted that though the response is accuarte in terms of content, contextual and cultural references were not captured when response is provided in English.

Our analysis also indicated that the chatbot tended to interpret questions literally, without considering the user's background or context.
For instance, Amar (a chatbot developer at Myna) shared his observation over a WhatsApp conversation---\textit{``baccha kyu nahi rukta'' has been interpreted as `why doesn't baby stop crying' when the actual intention of the user was to ask `why Can't I get pregnant' ''}. This gap between the user's intentions and how they framed their query could result in an incorrect response by the chatbot. Our focus groups with Myna revealed that in an early effort to address such issues, they tried fine-tuning using the chatbot by providing a developed question-answer dataset as training data. However, the poor performance of the model and the relative financial and computational cost led the team to continue to use the RAG approach and focus on improving intent detection and the knowledge base.

We also noticed that the chatbot did not always correctly understand that a question relating to socioeconomic status was being asked, misinterpreting the user's intent and providing an inaccurate response. For example, Table \ref{tab:emp} shows that the user query on \textit{``How important is} my \textit{job for family planning''} resulted in the chatbot responding how it can play an important role in family planning. 
This issue stemmed from an error in the translation module of the system, which translated the user query to \textit{``How important is }your\textit{ job for family planning''}. 
While the multi-stage system design that involved first improving grammar and then translating and generating text in English was effective in most cases, it also occasionally introduced other errors in the process.

To improve the system design, the software development team at Myna introduced a localization module in the system, which involved maintaining a dictionary where medical terms generated by the LLM were replaced with local terms instead. 
Beyond the use of medical terms, we also found that the word choice of the chatbot at times was not always appropriate, even if the chatbot had interpreted the user question accurately and the response as a whole was medically accurate.
This concern came up with respect to a chatbot response to a question on contraceptive sponges. 
The chatbot responded with: {\textit{``Garbhnirodhak sponge ek soft, disk-shaped, polyurethane foam ka tukda hota hai jo spermicide se bhara hota hai. Yeh bhi yoni mein daala jata hai aur yeh shukranuon ko rokne ke saath-saath unko marne ka kaam bhi karta hai.''}} (which translates to \textit{``A contraceptive sponge is a soft, disk-shaped piece of polyurethane foam that is filled with spermicide. It is also inserted into the vagina and works both by blocking sperm and by killing them.''}. 
Dr. Preeti shared that the chatbot should have used \textit{``Istemal''} (which translates to ``use'' in English) instead of \textit{``daala jata hai''} (which translates to ``is put''). The latter could be perceived as being too explicit.
% despite conveying similar meanings.

% % While the latter may be used occasionally 
% In general, we observed that the chatbot had not been designed to respond to social norms like greeting the user before starting a conversation. It would immediately start to respond to the user's prompt without distinguishing terms that are considered respectful or disrespectful in various cultures.
% \textbf{

% For example, when a user asked a question about family planning, the chatbot responded directly with `\textit{Nahi, family planning ke nirnay lene wale sirf education wale log nahi hote. Family planning har vyakti ke liye mahatvapurna hai, chahe ve shahar mein rahein ya gaon mein.}''. This response had relevant information but missed culturally appropriate greeting, which is essential in communication in many contexts.
%\subsection{Improving Accessibility for a Broad Audience}
\begin{table*}[!h]
  \centering
   \fontsize{7}{8}\selectfont
  \begin{tabular}{>{\raggedright\arraybackslash}p{1.5cm} p{2cm} p{2.5cm} p{8cm}}
    \toprule
   \textbf{Topic} & \textbf{Language} & \textbf{User Query} & \textbf{Chatbot Response} \\
        \midrule
 \multirow{2}{*}\chiadd{Health Literacy} &English (Translated) & Does Saheli tablet change the date of periods? &
Saheli, also known as Centchroman, is a non-steroidal oral contraceptive pill. Sometimes this can lead to changes in the menstrual cycle, due to which the dates of periods can change. This is because it can affect the hormones that regulate the menstrual cycle. If you notice any significant changes in your cycle or have any concerns, it is important to discuss this with your healthcare provider. Remember, every woman's body reacts differently to contraceptives, so it is beneficial to keep track of your cycle and note any changes. If you feel uncomfortable or have any questions, please consult a Myna's Telehealth doctor.\\
\cmidrule{2-4}
 &Hinglish & Saheli tablet se periods ka date badal jata hai kya? & Saheli, jo Centchroman ke naam se bhi jaani jaati hai, ek non-steroidal oral contraceptive pill hai.
Kabhi-kabhi yeh menstrual cycle mein parivartan laa sakta hai, jiske karan period ki dates mein
badlav ho sakta hai. Iska yeh wajah hai ki yeh hormones par asar daal sakta hai jo menstrual cycle ko
niyantrit karte hain. Agar aapko apne cycle mein koi bhi bada parivartan mehsoos ho ya koi bhi chinta
ho, toh isko healthcare provider ke saath discuss karna mahatvapurn hai. Yaad rakhein, har mahila ke
sharir ka prativarsh contraceptives ke prati alag prakar se react karta hai, isliye apne cycle ka dhyan
rakhna aur kisi bhi parivartan ko note karna faydemand hai. Agar aapko asamanata mehsoos ho rahi
hai ya aur koi sawaal ho, toh kripya Myna’s Telehealth ke doctor se paramarsh karein.\\

\midrule
 \multirow{2}{*} \chiadd{Sterilization}&English (Translated) & Can vasectomy fail after 3 years? & The chances of a vasectomy failing are very low, but technically, it is possible that a vasectomy may
fail even after several years. Its main function is recanalization, in which the separated parts of the
vasa deferens start reattaching themselves and thankfully beginning to meet the semen. However,
this is very rare, especially if the initial tests after vasectomy are not found successfully.
If you are concerned about conceiving after a vasectomy, it would be appropriate to have a semen
analysis done. This test will help you confirm whether the vasectomy is still effective.
Do you need any further information about the long-term effectiveness of vasectomy?\\
\cmidrule{2-4}
&Hinglish &Kya 3 sal bad purush nasbandi fail ho sakti hai?&Vasectomy ke fail hone ke chances bahut hi kam hote hain, lekin technically, yeh sambhav hai ki
vasectomy ke kai saal baad bhi fail ho sakta hai. Iska mukhya karan hota hai recanalization, jisme vasa
deferens ke kata hua hisse apne aap phir se jud jaate hain aur shukranu veerya mein milne lagte hain.
Lekin, aisa hona bahut hi durlabh hai, khaas taur par agar vasectomy ke baad ke shuruaati veerya
parikshan mein shukranu nahi paaye gaye ho.
Yadi aapko vasectomy ke baad garbh dharan ki chinta hai, to veerya parikshan (semen analysis)
karwana uchit hoga. Yeh jaanch aapko yeh pushti karne mein madad karegi ki vasectomy abhi bhi
prabhavi hai ya nahi.
Kya aapko vasectomy ke long-term prabhavi hone ya isse judi aur koi jaankari chahiye?\\
\bottomrule
  \end{tabular}
%    \vspace{10pt}
  \caption{\textbf{Understanding Information-seeking behaviors around SRH}. An instance of the chatbot's use of several complex medical terms and chatbot response to a question on the vasectomy.}
  \label{tab:literacy}
  \Description{Table shows an instance of the chatbot’s use of several complex medical terms and chatbot response to a question on vasectomy with both original Hinglish and English text. }
\end{table*} 

\subsubsection{Being responsive to a range of health and digital literacy levels}
From our interviews and by analyzing log data, we found that health and digital literacy played an important role in how the users understood the responses and interacted with the chatbot. 
Healthcare professionals at Myna expressed a preference for audio responses over textual responses considering users' digital literacy. However, several issues were observed with the audio output of the bot. Words that are commonly not found in English, such as the name of the NGO itself (which was in Hindi), were pronounced incorrectly. 
Some complex medical terms were omitted entirely in the audio responses such as ``Etonogestrel''.
We also observed that chatbot responses frequently consisted of complex medical terms.
The user interaction in Table \ref{tab:literacy}, \chiadd{under the health literacy topic}, shows how the chatbot used complex medical terms like ``non-steroidal'', ``menstrual cycle'', ``hormones'', ``contraceptives''.
Dr. Preeti also shared this concern, and observed that the chatbot sometimes used very formal words in Hindi that are not commonly used and understood by all the users:
% There are instances where the chatbot used English words in its responses even when the user’s query is entirely in Hindi
 % and a lower level of comprehension among users. 

\begin{quote}
``like doctor is called `swaasthya seva pradaata', expert advice is called `peshevar salah', medically accurate is called `chikitsakiya tareeke'. these would not be understandable by people''. 

—Dr. Preeti (Health Care Professional)
\end{quote} 

Another example in Table \ref{tab:literacy} \chiadd{on sterilization} shows how the chatbot has used a different set of medical terms in its response, instead of using the same local terms that the user employed in their question. 
This can impact the user's understanding of the complete response, as they may not recognize the terms used. 
% Initially, chatbot was not able to simplify complex health information for users who may have limited education or understanding of medical concepts. 
Over time, with the introduction of local words for medical terms in the dictionary and the expansion of the knowledge base by the Myna team, the chatbot was able to replace complex terms with simple terms improving comprehensibility.

\subsection{Taking Social and Cultural Norms into Account}
% \chiadd{There is also the notion of what constitutes a misconception versus a belief. Misconceptions, often rooted in factual inaccuracies, and beliefs, shaped by personal, cultural, or religious values, may require either correction or alignment, depending on the context. Our findings highlight the complexity of navigating these dynamics.} 
We now focus on how the chatbot engaged with users' social and cultural contexts when responding to queries. We discuss the need for balancing medical accuracy with respect for cultural beliefs, when providing information on sensitive topics like family planning and sexual health. 

\subsubsection{Creating a space to discuss cultural taboos}
Our focus groups with the \chiadd{women from the community}\chirm{workers} revealed that the chatbot offered a channel to ask questions that they might not have asked at all otherwise. One worker shared that, \textit{``Before this work, I had not thought about these questions. But because of this job, I started to try to come up with questions and ask questions from my sister, in-laws, and neighbors, and look online for ideas of questions to ask. It was very difficult to come up with them.''}
% In many conservative cultures, women prefer to see female doctors for sensitive health issues because they feel more comfortable discussing SRH topics with them. 
% This preference is often due to cultural taboo around modesty and potential belief that it is inappropriate to discuss SRH issues with someone of the opposite gender. The comfort associated with preserving modesty and adhering to cultural expectations of privacy significantly influences the choice of a female doctor over a male doctor, a sentiment that was echoed by 
Dr. Pooja also reflected on her years of experience working with marginalized women in the community---\textit{``People like to talk to female doctors because they can't talk to male doctors more openly but can talk to these apps''}. She continued to add, \textit{``Young married women aged 18 to 30 said they want to speak up about SRH, but it's very difficult for them to get those words. They feel uncomfortable. Few of them mentioned that they would only ask their husbands about such matters, as talking to neighbors was difficult due to concerns about privacy, shame, and stigma around it''.} The analysis of the questions asked presented in Table \ref{tab:type} earlier, also indicates that the design of the chatbot was largely effective in eliciting questions on deeply personal, sensitive, and sometimes controversial topics.

%significant challenges faced by users, particularly young married women aged 18 to 30, when discussing sensitive topics like family planning and reproductive health. Many users experience shame, discomfort, and fear of judgment, leading them to avoid conversations about these issues, even with close family members including mothers and sisters. This cultural taboo and desire for privacy often result in users limiting their discussions to only their husbands, in some cases, not seeking advice at all. 
Our analysis showed that the chatbot maintained a neutral tone when addressing queries associated with cultural taboo (see Table \ref{tab:Baby Sex} in Appendix), which helped in creating a comfortable environment for users, encouraging them to discuss sensitive topics that might otherwise be avoided in real-world interactions. This also allowed users to seek information or advice from the chatbot without fear of judgment. For example, one of the user questions was---\textit{``I don’t know how to have sex. I have just recently gotten married''}. This indicates the users' comfort level with the chatbot to seek information on sensitive topics. 

\subsubsection{Correcting misconceptions while respecting cultural beliefs} 
Our findings indicated that religious and community beliefs could shape health behaviors. 
We identified queries around beliefs that stemmed from misconceptions, and both harmful and non-harmful community practices. For instance, a user asked a question on preferential sex selection, as seen in Table \ref{tab:Baby Sex}. Prior work by Neogi et al. has examined how Indian women use certain medical and non-medical techniques with the belief that these methods could influence the sex of their child \cite{Neogi2015ConsumptionOI}. The chatbot only provided a response that was medically accurate and scientific. It did not recommend any unsafe and unproven techniques to further boost the user's beliefs.

\begin{table*}[h]
  \centering
     \fontsize{7}{8}\selectfont
  \begin{tabular}{>{\raggedright\arraybackslash}p{1.5cm}p{2.5cm} p{10cm}}
    \toprule
  \textbf{Topic} & \textbf{User Query} & \textbf{Chatbot Response} \\
        \midrule
   \chiadd{Misconceptions and Gender Norms} & In family planning, I only want a boy, Are there any methods for that??
 &
 There is no guaranteed way to choose the gender of a baby. The main objective of family planning is to control pregnancy, not to select a child's gender. Some people try different methods, such as timing and specific techniques, to predict whether they will have a boy or a girl, but these methods are not scientifically proven and have no reliable basis. If you are considering family planning options, it is important to focus on methods that are best suited for you and your family. If you would like more information or need any specific advice, please consult a doctor with Myna's Telehealth. \\
  \midrule
  \chiadd{Cultural taboo} & Why shouldn't the word 'sex' be used in front of family when discussing family planning? &
In every family and community, communication styles vary. In some families, the term 'sex' is not used openly because it may be uncomfortable for them. This can be due to social or cultural beliefs. When it comes to family planning, people often discuss the topic in a more considerate manner, using terms like 'making relations', 'family planning', or 'planning for children'. It is important to respect the feelings of your family members when discussing this topic and to choose words that are comfortable for everyone. This helps in having open and meaningful conversations without making anyone feel uncomfortable.\\
\midrule
 \chiadd{Dietary recommendations} & What to eat to heal family planning sterilization wounds &
   Eating a nutritious diet is essential for proper wound healing after sterilization. Here are some dietary recommendations that can help in recovery: 
1. Protein rich diet: Protein is important for wound healing. Eat protein-rich foods like dal, chana, soybean, paneer, curd, eggs, chicken, and fish.
2. Vitamin C and Zinc: Both these dietary elements help in wound healing. Oranges, lemons, kiwi, plums, bell peppers, walnuts, and seeds should be consumed in the diet.
3. Green Vegetables: Spinach, fenugreek, sarson, and other green vegetables are rich in vitamins and minerals that help in tissue repair.
4. Fruits: Fresh fruits like apple, banana, papaya, and berries are rich in antioxidants which keep the body healthy.
5. Adequate water: Drink adequate amount of water because hydration is necessary for wound healing.
6. Whole Grains: Whole grains like wheat, barley, and oats are rich in fiber which keeps the digestive system healthy.
In addition to this diet, diets containing alcohol, caffeine, and other sugars should be avoided as they can slow down the wound healing process. Do you need any more suggestions or information? \\
\midrule
\chiadd{Religion} & What to do if surgery is prohibited in religion? &
    If your religion or beliefs are against operations or surgical procedures, you should consider some non-surgical birth control methods. There are some non-surgical methods:
1. Condoms: This is an effective method that protects against pregnancy and sexually transmitted infections (STIs).
2. Oral contraceptive pills: Taken daily by women to prevent pregnancy.
3. Vaginal rings: These are inserted into the vagina for a month at a time.
4. Contraceptive patches: These are patches applied on the skin which have to be changed weekly.
5. Natural family planning methods: Like ovulation tracking and withdrawal method.
Any of these options may be right for you. But, before using them, it is important to understand that some methods, such as natural family planning, may be less effective. 
You should seek guidance from your religious leader or community as well as a doctor. You can also contact ``Myna's Telehealth'' where you can consult a doctor who can give you the right guidance taking into account your religious beliefs.\\
    \bottomrule
  \end{tabular}
%%    \vspace{10pt}
\caption{\textbf{Taking Social and Cultural Norms into Account.} Chatbot response to questions on cultural beliefs, taboo, food recommendations and role of religion. Find the original
Hinglish text in Table \ref{tab:stigma} (in the Appendix)}
    \label{tab:Baby Sex}
    \Description{Chatbot response to user questions on cultural beliefs, taboos,
food recommendations and the role of religion are presented in the table with English text. The original Hinglish text is referenced to Table \ref{tab:stigma}}
\end{table*} 

In one instance, a user asked if condoms cause HIV. The chatbot responded: \textit{``Yes, it is important to use a condom every time you have sex to protect against HIV and other sexually transmitted diseases.''}. Though this was scientifically accurate, it did not address the underlying misconception of the user that condoms do not cause HIV and are, in fact, a preventive measure. This may have been a missed opportunity to correct misinformation. Identifying when a user's question reveals a misconception is critical to address any harmful myths that may otherwise lead to uninformed health decisions.

We also found cultural beliefs come up in other ways, such as with dietary practices. However, the chatbot did not consider the specific dietary preferences of individuals when responding to questions on this topic. The user interaction on food recommendations listed in Table \ref{tab:Baby Sex} shows that while the recommended food choices were Indian, they were quite broad and did not consider the user's specific dietary needs or preferences and affordability. Given India's diversity, dietary choices are often influenced by factors such as religion, region, festivals, and socioeconomic status. 

\subsubsection{\chiadd{Providing relevant healthcare service suggestions} \chirm{Considering the role of religion}}
We also found that the sociocultural context could shape what healthcare services were considered acceptable by users. We present two examples below where this was apparent---in relation to religious beliefs and socioeconomic status. 

In the example presented in Table \ref{tab:Baby Sex}, the user asks a question about what to do if surgery (sterilization) is prohibited in one's religion. Rather than rebutting this claim, the chatbot offers alternative non-surgical methods and also suggests seeking guidance from a religious leader as well as a doctor.
% we observed that the chatbot placed religion and the benefit of the user at the same level, which reflects a culturally sensitive approach. However, it does not establish a hierarchy that prioritizes safety over religious considerations. 
In another user query on the acceptability of family planning in the Muslim community, the chatbot responded by saying:{\textit{``Absolutely, family planning is necessary in every social community, be it Muslim community or any
other. This is something every family should think about, because it affects not only your health,
but also your children’s care and financial status. There are some easy and permissible methods for
family planning in Islam, such as temporary methods such as condoms or permanent methods such as
tubectomy or vasectomy. What do you think, do you need more information about family planning?.''}}
% (which translates to \textit{``Yes, it is important to use a condom every time you have sex to protect against HIV and other sexually transmitted diseases.''}.
While the response did not get into the specifics of which religious scriptures indicate whether or not family planning is permissible, it advocates more generically for family planning in every community. 
% nstead, it provided a more generic response.
Once again the response suggests talking to a religious or community leader.
While the balance between sharing medical information and religious beliefs was handled diplomatically in the two examples, there are situations where conflicts could emerge.
% this could present . 
For instance, in the case that a religious practice or belief is harmful or medically inaccurate, then the safety of the user is critical, and medical advice should be emphasized.

Our findings showed that the chatbot did not always take into account the socioeconomic status of participants, which could shape access to healthcare services. 
Some of the services or practices  (e.g. IVF) suggested by the chatbot could be quite expensive and impractical for participants.
% Despite providing a factual response, the chatbot did not address the underlying concern of access to medical services based on income. 
% Access to medical services is highly dependent on income and employment status. 
One user specifically asked a question on whether family planning can be done if there is no money. The chatbot responded by saying---\textit{``Of course, it does not matter whether you have money or not for family planning. Many measures can be beneficial for you.''} However, it did not provide tailored information on free or low-cost family planning resources available through government or non-profit initiatives. % \cite{article}

\subsection{Responding to Social Dynamics}
We now address how the chatbot responded to queries involving social dynamics. We highlight the challenges in interpreting the social and emotional contexts of user queries, such as family pressure around pregnancy or the complexities of marital relationships.

\subsubsection{The role of legal implications}
The chatbot was inconsistent in responding to legal aspects of user queries, which mostly related to the appropriate age for sex and marriage. 
It generally advised users to follow legal procedures, but did not always offer specific information on legalities applicable to that context. 
For instance, Table \ref{tab:planning} presents a response to the user question---\textit{``What is the proper age for having sex for the first time.''} The chatbot responds by saying that the legal age varies in every country and is between 16 and 18 years, rather than providing the legal age specific to the user context (the age of consent is 18 years in India). 
In another example, however, the chatbot did not address the legal implications related to sex at all.
Table \ref{tab:planning} presents an interaction about whether a 15-year-old can think about family planning, but does not discuss the legal implications of having sex at that age at all.
This was a critical gap that the Myna team attempted to address by improving intent detection and updating the knowledge base with more legal information specific to the study context.
% The inconsistency in the chatbot's ability to recognize and apply context specific legal information should be emphasized in order to maintain system's reliability and trust among users.

\begin{table*}[h]
\centering
      \fontsize{7}{8}\selectfont
  \begin{tabular}{>{\raggedright\arraybackslash}p{1.5cm} p{2.5cm} p{10cm}}
    \toprule
      \textbf{Topic} & \textbf{User Query} & \textbf{Chatbot Response} \\
      \midrule
     \chiadd{Age of sexual maturity} & What is the proper age to have sex for the first time? &
The concept of `proper age' for sex depends on legal and emotional maturity. In every country, the legal age of having sex (also known as the `age of consent') varies, and is usually between 16 and 18 years. This legal age is required to ensure that a person is emotionally and physically mature enough to understand his/her decisions and handle their consequences. But, merely being of legal age is not enough. It is also important that the person is emotionally ready, has the right information, and is ready to have a healthy and balanced relationship with his/her partner. It is also important to understand the importance of safe sex and consent. Therefore, the `right time' to have sex for the first time depends on the person's own thoughts, feelings, and circumstances. If someone needs more help or information in this matter, it would be wise to talk to a trusted doctor or healthcare provider.\\
\midrule 
    \chiadd{Family planning} & Can a 15 year old girl do family planning?
 &
Yes, a 15 year old girl can think about family planning, but in this age, mostly the focus will be on education and personal development. If family planning is needed, you should first consider non-invasive methods such as barrier methods (condom use) or oral contraceptives (pills). Both of these methods are safe and reversible, meaning you can stop using them whenever you want. But, before taking any decision, it is important to consult a specialist or doctor of Myna's Telehealth. His guidance will guide you in the right direction.\\
  \midrule
   \chiadd{Divorce} & If I don’t stay with my husband after marriage, is it right to get a
divorce? &
  If you are no longer able to get along with your husband after marriage and you feel that there are problems between you, then it is important to first understand what the problem is. Every relationship is different and sometimes it can be difficult to understand or find a solution. But, it is also important that both of you talk to each other openly and try to find a solution to the problem. If you feel that you have tried everything and still cannot find a solution, then you should listen to your heart. This is a very big decision, so take time to think about it and also consult a specialist or counselor if necessary. Your happiness and health come first. \\
   \midrule
   \chiadd{Gender roles} & Why don't men do vasectomy? &
Male sterilization, or vasectomy, is a permanent contraceptive procedure that is very safe and effective. But, sometimes men do not undergo this procedure due to many reasons:

1. Lack of information: Many men do not have the correct knowledge about this procedure.
2. Misconceptions: Some men believe that sterilization will affect their sexual strength or masculinity, which is incorrect.
3. Social Pressure: In some communities, there is a belief that sterilization is only meant for women.
4. Fear: The fear of surgery or medical procedures can also be a factor.

Vasectomy does not reduce a man's sexual ability nor does it cause any changes in male hormones. This is a safe and less painful procedure. If you or someone you know is interested and would like more information, please contact your local health center or doctor. Would you like to ask anything else about this?\\
  \bottomrule
  \end{tabular}
%    \vspace{10pt}
    \caption{\textbf{Responding to Social Dynamics.} Chatbot responses to questions on legal implications, and gender roles. Find the original Hinglish text in Table \ref{tab:mar} (in the Appendix)}
      \label{tab:planning}
      \Description{Chatbot response to user questions on legal implications, and gender roles is presented in the table with English text. The original Hinglish text is referenced to Table \ref{tab:mar}}
\end{table*}

\subsubsection{Considering family and partner dynamics} 
We observed that the system currently is not able to understand the familial context in the user prompt. It focuses solely on the medical terms present in the prompt without heeding attention to the family context. From a WhatsApp chat conversation, we noted that there was a user query---\textit{``I am newly married, I don't want to get pregnant yet but my family is pressuring me to have children, so what should I do?''} The chatbot responded with information on methods of preventing pregnancy rather than addressing the user’s concern about handling family pressure regarding pregnancy. 
% The response should have been more aligned with providing guidance on managing family expectations and addressing emotional aspects related to the situation. 

%\subsubsection{Navigating relationship with the partner}
Our analysis also showed a potential limitation of the chatbot in providing personalized advice on important life decisions. In the case of a user question about divorce (Table \ref{tab:planning}), the chatbot ignored the context provided by the user in the query that the user's husband is not living with her and provided a generic response recommending improved communication or counseling. Counseling might not be accessible, affordable, or culturally acceptable in this context.
By not asking follow-up questions, the chatbot missed the opportunity to get more information on the dynamic with the partner and potentially provide more emotional support. 
We also note that this question went beyond the focus on SRH and focused on well-being in a relationship more broadly, pointing to other potential gaps in information access and socioemotional support among participants.
% Evaluating the specifics of the user context should be emphasized, considering users might make informed decisions based on the response. 
\subsubsection{Dealing with gender bias} 
Our analysis uncovered that the chatbot did not reinforce traditional gender roles in the home and around family planning.
% Our analysis uncovered a few inconsistencies in how the chatbot when responded to gender-related queries.
% First, we observed that the chatbot gave generic responses to topics like family planning without accounting for cultural variations in gender roles. 
For example, in one of the user's queries about the women's role in family planning, the response of the chatbot was: \textit{``No, the responsibility of family planning does not rest solely on women. It depends on the understanding and cooperation of both partners. Men can also contribute to family planning, such as by using
condoms, opting for vasectomy (male sterilization), or supporting their partner’s family planning
decisions. Family planning should be a shared decision, in which both partners participate and agree.
Do you need any further help on this topic?''}. 
% It gave medically accurate information but overlooked the cultural context in which gender roles may have an impact on family planning decisions.
This approach was by design as a result of Myna's focus on supporting women's agency and empowerment in healthcare decisions, as an organization. 

% \textbf{Further analysis showed that it maintained gender neutrality in responding to gender-related queries. It avoided making any assumptions and responded with consideration for all genders. Despite lacking gender specificity in some responses, 
This extended to also being able to respond to questions about why gendered expectations exist around reproductive health in the first place.
For example, a user asked a query about why men do not undergo vasectomy. The chatbot was able to acknowledge societal pressures and misconceptions around gender roles, particularly when discussing topics like vasectomy. It acknowledged social pressure as a possible reason why some men might hesitate to undergo surgery (see Table \ref{tab:planning}), and emphasized that ``manliness'' does not reduce as a result of undergoing a vasectomy. 
 % These findings indicate a gap in the system's ability to respond based on culturally specific gender norms.

%% file: discussion.tex
\section{DISCUSSION}
The results of our study underscore the importance of cultural sensitivity when designing LLM interventions, particularly for healthcare and stigmatized topics. While LLMs offer promising opportunities to improve access to health information, especially in resource-constrained settings, their effectiveness depends on their ability to respond to complex cultural contexts. In this section, we first consider what constitutes culture in the LLM context, \RD{and how our understanding relates to approaches in prior work.} We then present a framework for the design of culturally sensitive LLM-based chatbots, \RD{also taking inspiration from related chatbot research}, and reflect on the challenges and complexities inherent in this process. 
Through this discussion, we aim to shed light on key considerations for designing AI systems that respect cultural norms while delivering medically accurate and actionable health information. 

%\RD{While existing chatbot applications have focused on integrating certain cultural beliefs within the community, they often remain limited in their ability to adapt to user-specific beliefs and evolving conversational contexts. Our approach refines prior work by leveraging a trained LLM and unpredictable real-world conversations ensuring that interactions remain contextually relevant. Departing from the previous approaches \cite{yadav_feedpal_2019, wang2022artificial}, our framework is designed to be adaptive,  modifying its responses in real time based on user input and cultural nuances, making it a more practical solution for diverse healthcare interactions. }

\subsection{What Constitutes ``Culture'' in LLM Development?}
Our focus on culture was driven by the motivation to address sensitive health topics in a manner that respects the norms of the community. 
While medical accuracy was considered by the Myna team to be the most critical criteria, cultural relevance was seen as being equally important to ensure that the information provided was relevant to the needs of community members and accepted by them despite the taboo nature of the topic.
% When we initially, engaged in. 
However, integrating cultural relevance was a non-trivial task. Our findings illustrate how culture is deeply intertwined with other context dimensions, such as social norms, accessibility, and legal considerations.
Though we presented our findings in neat categories, many of the user queries touched on multiple of these elements.
For example, cultural beliefs regarding the appropriate age for family planning (Table \ref{tab:mar}) can be closely linked to regional laws governing the minimum age for consent and marriage, and societal expectations around gender roles. In such cases, it becomes difficult to isolate multiple elements, as they mutually shape individuals' perceptions and decisions. 

Setting aside these complexities for a moment, let us revisit the definition of culture that we started with. We initially relied on the definition of culture from health communication literature as \textit{``shared values, norms, codes, roles, and assumptions that shape a group's beliefs, attitudes, and behavior through their interactions in and with their environments''} ~\cite{griffith2024cultural}. 
In a recent review of literature on LLMs by, Adilazuarda et al. present a ``taxonomy of culture'' based on proxies that they identified in LLM literature  \cite{adilazuarda2024measuringmodelingculturellms}. In contrast, we take a bottom-up approach to categorizing the role of culture in LLMs, by studying how users asked questions in a specific context.
We offer a definition of a \textit{culturally sensitive} LLM as---\textit{an LLM (or LLM-based service) that provides verified \chiadd{and verifiable} information while considering the user's beliefs, attitudes, and behavior, shaped by \chirm{their cultural backgrounds} \chiadd{the shared values, norms, codes, roles, and assumptions prevalent in their community.}}
% The challenge with integrating culture into LLM-based chatbots, even more so than in the case of other digital technologies, is that they attempt to reach a wide audience but have to respond to the particularities of individual users.
We deliberately consider the impact on the individual user \chiadd {(rather than group messaging which is common in health communications),} because many LLM-based applications support highly personalized one-to-one conversations.

%\RD{Previous work primarily focused on regional and community-level elements, Our framework extends beyond these limitations in a significant way by incorporating societal and individual considerations.}

\begin{figure}[h]
  \centering
\includegraphics[width=0.5\textwidth]{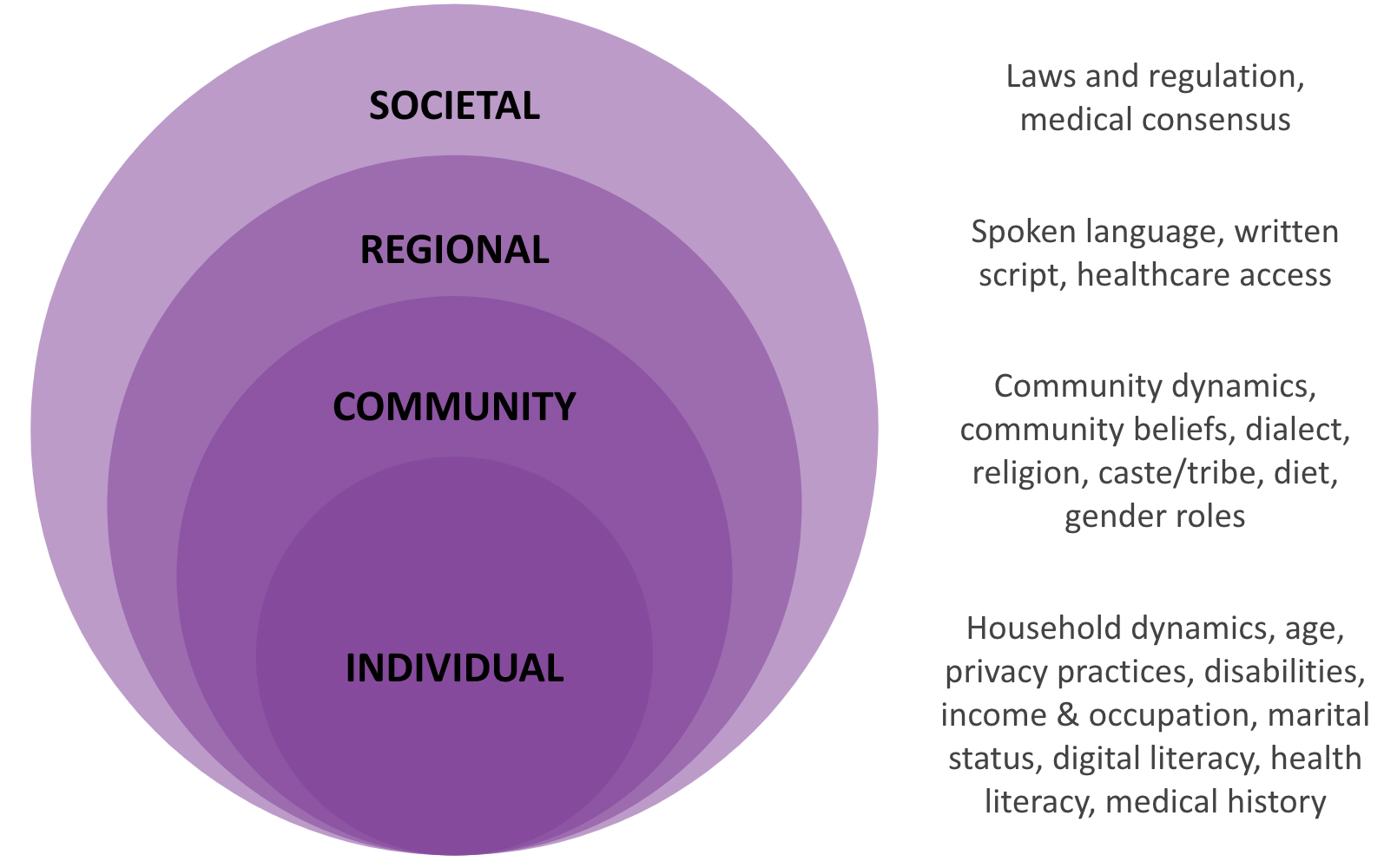} 
  \caption{\textbf{The four context layers that shape cultural relevance of LLM-generated text}.}
  \label{fig:contextlayers}
  \Description{Figure that shows four context layers that shape the cultural relevance of LLM-generated text starting with the Individual layer from the bottom followed by Community, Regional, and Societal layers.}
\end{figure}

This leads us to propose a multi-dimensional approach to understanding the role of culture in the design of LLM-based interventions. We categorize context as four layers, each enveloping the other--- societal, regional, community, and individual. 
% Each of these dimensions entails multiple elements of context that operate at that level. 
Culture is most strongly linked to the three high-level layers (society, region, and community), which shape individual beliefs, attitudes, and behaviors (see Figure \ref{fig:contextlayers}).
Each of these layers is further broken down into multiple dimensions; most though not all of them are tied to culture. These dimensions emerged from our inductive coding process.
 % are shapedinfluence what happens at the individual level
We present the detailed framework next.

%Structured knowledge bases in chatbots can improve this by providing more relevant and context-aware responses, enabling users to navigate information more effectively. Enhanced interaction capabilities through knowledge organization can also empower users to engage more meaningfully with chatbots.

% This balancing act underscores the complexity of addressing healthcare in diverse cultural environments, where societal, regional, and individual factors are interdependent.

% Medical accuracy, for instance, must be communicated in a way that aligns with cultural beliefs without compromising factual integrity. 
% Medical accuracy.

\begin{table*}[h]
% \fontsize{6}{5}\selectfont
% \footnote
\centering
\small
  \begin{tabular}{>{\raggedright\arraybackslash}p{1.2cm} p{2.6cm} p{4.3cm} p{6cm}}
% \begin{tabular}{l l l l}
\toprule
\textbf{Context Layers} &\textbf{Dimension} & \textbf{Example of Relevance to SRH or LLM Choices} & \textbf{Implications for Design} \\
\toprule
Societal  & Medical Consensus & Medication approved for use by regulatory bodies; consensus on how to prevent spread of HIV & Include in Knowledge Base (KB)\\
& Laws and Regulations & Legal age of consent for sex & Include in KB \\
 %  & &   & \\ 
 \hline

Regional & Spoken Language & Use of Hindi, Marathi, and Urdu widely in Mumbai & Pick LLM model or translation service to support that language \\
  % & & &  \\ 
 & Written Script & Use of transliterated Hindi widely in WhatsApp communications & Gather and study examples of everyday chat communications; consider fine-tuning with this data if translation does not work well or existing LLMs for that language perform poorly\\
% & &  &   \\
% & & &   \\
% & & &  \\
 & Healthcare Access & Lack of affordable clinics in the area & Suggesting accessible actions such as offering teleconsulation or referring to free or local services \\
% & & &  \\
% & & &  \\
  \hline
  
Community & Community Dynamics & Taking a neighbor's health advice on how to reduce pain during periods & Update KB to recognize importance of community,  and encourage consulting a community leader for advice, while prioritizing verified medical information \\
   & Community Beliefs & Lack of belief in vaccinations & Counter misconceptions, counter harmful practices, maintain a neutral tone if it is a benign practice  \\  
  %  % & & & any harmful practices, \\
  % & & &  \\
 & Religion & Belief that sterilization is not allowed in their religion & Update KB to recognize religious and communal beliefs, encourage talking to a religious/community \\
 & Caste \& Tribe & Belief that a woman is impure while menstruating &  leader, and offer alternatives if a recommendation goes against their beliefs \\
     % &  &  &  if they believe that a certain recommendation 
  % &  & eligibility for government schemes & \\  
 & Dialect & Specific manner of speaking or writing & Dictionary to swap out with words used locally \\
& Gender Roles & Limited mobility of women shaping healthcare access & Prompt asks LLM to acknowledge dominant gender roles, but also that they can change. Center  women's agency and offer strategies to negotiate power over one's health and in their relationships\\
 & Diet & Dietary preferences & Update KB with recommendations based on local dietary practices \\
   \hline
   
Individual & Household Dynamics & Discomfort talking about sex with partner & Include suggestions for navigating family dynamics in KB \\
 & Privacy Practices & Woman shares device with her son & Design the LLM prompt to generate text with a tone that is respectful and formal, similar to a medical professional\\
 & Age & Experiencing menopausal symptoms & Ask follow-up questions on age when relevant to a health symptom, and factor age in prompt when brought up by user \\
 & Income \& Occupation & Stigma around accessing free government services & Consider when suggesting actions to take such as offering teleconsulation or referrals \\
 & Marital Status & Taboo on using contraceptives (having sex) before marriage & Update KB to provide information while acknowledging taboos \\
 % &  &   &\\ 
%  & Occupation & Works in a labor-intensive job and cannot rest & \\
% & &   during pregnancy  & \\
 % & Education & Learned reproductive health in 10th grade &\\
 & Digital Literacy & Discomfort with typing resulting in grammatical errors & Do grammar correction, offer voice capabilities \\
 & Health Literacy & Unfamiliar with a medical term & Continually update dictionary and KB with simple language\\
 & Medical History & Having a prior miscarriage; or experiences with heavy and painful periods & Ask follow-up questions on medical history relevant to a health symptom, and include medical history in the prompt when brought up by the user \\
 & (Dis)Abilities & Mobility impacting healthcare access & Offer accessible services like teleconsultation \\ 
\bottomrule
  \end{tabular}
    \vspace{-5pt}
 \caption{\textbf{Framework for Culturally Sensitive Design of LLMs in Healthcare.} The acronym KB refers to the Knowledge Base.}
  \label{tab:framework}
  \Description{Framework designed for culturally Sensitive Design of LLMs in Healthcare is presented with context layers, dimensions for each layer, examples of relevance to SRH, and implications of design. The acronym KB used in the framework refers to the Knowledge Base. }
  \end{table*}

\subsection{Framework for the Design of ``Culturally Sensitive'' LLM Interventions}
% \RD{Yadav et al. and Wang et al.'s work align with Resnicow et al.'s primary dimensions of cultural sensitivity \cite{resnicow1999cultural}. At the surface structure \cite{resnicow1999cultural}, they supported code-mixing and voice input, enhancing accessibility for users. More significantly, their approach considered elements of deep structure \cite{resnicow1999cultural} such as social influences, particularly the role of mothers-in-law in shaping maternal health decisions \cite{yadav_feedpal_2019}, and addressed social stigma surrounding SRH discussions among youth \cite{wang2022artificial}. 
% In comparison to SnehAI and FeedPal, which are not LLM-based, our approach refines their work by leveraging a trained LLM and unpredictable real-world conversations. While we take inspiration from their work, our findings prompted us to emphasize cultural beliefs more explicitly, recognizing the need for a more comprehensive framework that also accounts to user-specific beliefs and evolving conversational contexts.
% }

The use of LLMs for healthcare applications calls for a structured approach to ensure that culture is considered in design.
Based on our analysis, we developed a framework (Table \ref{tab:framework}), which summarizes the key layers influencing interactions in healthcare contexts. This framework identifies societal, regional, community, and individual layers that we propose be considered when designing culturally sensitive LLM interventions. Each of these has further been broken down into dimensions associated with them, such as laws and regulations, dialects, community beliefs, medical histories, and more.
We envision this framework as a starting point to help shape both the design and evaluation of LLM-based healthcare applications.

The framework aims to support the design of applications that are flexible enough to accommodate cultural dimensions across these various layers.
% , ensuring that users receive health advice that is not only medically sound but also culturally appropriate.
For instance, considering the \textit{societal} level, the system should always provide medical information that adheres to local laws (such as the age of consent) and current medical consensus. At the \textit{regional} level, it should adapt to spoken languages and dialects, and written script. \RD{Prior studies in other domains have also emphasized the significance of regional languages and local dialects as cultural indicators \cite{li2024should}. Within healthcare (though not using LLMs), Wang et al. and Yadav et al. have studied code-mixing and the importance of language adaptation in chatbot design \cite{wang2022artificial,yadav_feedpal_2019}.}
Additionally, our framework considers \textit{community} dimensions, such as the influence of religious beliefs or community dynamics on health behaviors. \RD{In one study, Rahman et al. also considered the role of religious beliefs in an adolescent sexual health education chatbot \cite{rahman_adolescentbot_2021}.} \chiadd{Myna's own chatbot design encourages users to consult community leaders or healthcare providers, thus taking into account how users’ social networks influence chatbot interactions and how they verify and act on information presented by a chatbot.} 
At the \textit{individual} level, understanding factors such as household dynamics, literacy, and medical history is crucial for personalized and effective health advice. \RD{Other studies have considered this aspect, such as the role of mothers-in-law in shaping health decisions in the context of a breastfeeding education chatbot \cite{yadav_feedpal_2019}}.
\RD{A study by Jo et al. on a voice-based LLM agent to address social isolation among older adults also illustrated the role of age, social networks, and digital literacies \cite{jo2023understanding}. 
% While not explicitly noted as a cultural element, their findings highlight how older users valued LLM-driven chatbots for emotional support \cite{jo2023understanding}. 
Their findings highlighted how users valued emotional support, but were also disappointed with the lack of individual personalization \cite{jo2023understanding}.} %particularly the chatbot's inability to remember their medical history.}

\chiadd{Our framework highlights that integrating cultural sensitivity may entail one or more of the following: the choice of the LLM or translation service in the first place, the choice between using pre-trained or fine-tuned models, integration of RAG to ensure accuracy in preserving cultural context while reducing hallucinations, updating the knowledge base with locally generated content and in response to new usage data, prompt tuning and including conversational history where relevant (with the prompt explicitly stating the specific voice or tone the LLM response will take, the cultural context to integrate, integration of community support, legal aspects, and more), and including a dictionary which can swap our text with locally-relevant terms. 
In addressing cultural sensitivity, it is also crucial to differentiate between aspects that can be integrated during the development process (including updating the knowledge base) and adopting a comprehensive, process-driven approach overall. While updating the knowledge base is an important step, it represents only one part of a broader, methodology required to achieve cultural sensitivity effectively. Enhancing the knowledge base with new information is insufficient because cultural contexts are deeply rooted in dynamic and subjective elements such as shared values, norms, codes, roles, assumptions, beliefs, attitudes, and behaviors.}

\chiadd{To navigate these complexities, we propose a scalable and generalizable \textit{methodology and development process} that prioritizes adaptability and community engagement (see Figure \ref{fig:methods} in the Methods section). 
We emphasize that cultural sensitivity is a \textit{spectrum}. We can never claim that we have achieved cultural sensitivity, especially given that culture itself can be dynamic, particularly in urban and increasingly online environments. We can only target doing better.
Our approach hence allows for iteratively integrating cultural elements into the development process, ensuring they are not static but evolve as the communities they serve evolve.
% By focusing on scalable processes rather than fixed outcomes, this methodology supports the integration of cultural sensitivity that can be applied across diverse populations.
Our outlined methodology for developing culturally sensitive chatbots consists of the following steps (Figure \ref{fig:methods}): (1) collaborate with community members and/or a local community organization to initially gather insights into their health concerns and identify key values and cultural elements, (2) work alongside the community and local healthcare professionals to co-develop a database of question-answer pairs to use for testing performance of pretrained LLM models and/or fine-tuning (and offer training when appropriate to enable this), (3) conduct iterative testing with local healthcare professionals and community members, and (4) monitor usage and incorporating user feedback into iterative improvements.
This approach underscores that cultural sensitivity is not a one-and-done goal but a continuous effort that evolves with a community's needs. By focusing on the ``how'' rather than just the ``what,'' this methodology hopes to integrate cultural sensitivity in a way that is both flexible and sustainable.}

\subsection{Complexities in Designing for Cultural Sensitivity}
Designing a chatbot to navigate cultural sensitivity presents significant challenges, particularly when even humans struggle to interpret and address sensitive cultural issues. The framework above aims to surface and present mechanisms to address these challenges. 
For example, terms like ``abortion'' versus ``miscarriage'' or ``family planning'' may have different connotations in the local context, leading to confusion or misunderstandings. The term ``abortion'' was used to refer to both miscarriages and abortions in our study context. On one hand, this has reduced the stigma around choosing to have an abortion. 
On the other hand, it has made it difficult to understand user context and provide appropriate advice based on whether someone has had an abortion or a miscarriage. 
Another major challenge lies in the language used by users, who may struggle to articulate their health concerns in a way the chatbot can understand. We noted how terms like ``cyst'' or ``Copper-T'' may be miscommunicated or misunderstood (e.g. leading them to be typed as ``cest,'' or ``cooperate T'' or ``copper tee'' respectively) due to limited familiarity with English and literacy levels. This linguistic complexity highlights the need for continuous updates to the chatbot's dictionary and prompt crafting, allowing it to better interpret emerging language patterns and provide more accurate responses.

Focusing on medical accuracy and cultural relevance at the same time can also lead to conflicts with dominant cultural, religious, or community health practices. We envision such situations being resolved as follows. If the health practice is benign and there is no documented evidence that it is not harmful, then the chatbot does not question the practice. If the practice is harmful, then medically accurate information is presented, without denigrating the practice but presenting evidence. If the health practice does not conflict with medical consensus, then the link to the community practices only helps to reinforce uptake. In all cases, we acknowledge the user's concern and encourage them to talk to their community leader and a doctor who can help them resolve the conflict better.

The notion of medical consensus in itself can be complicated in some contexts. Within India, not only are there competing community knowledges but also multiple complementary and alternative medicine systems. These are supported by the Indian government under the Ministry of AYUSH---Ayurveda, Yoga and Naturopathy, Unani, Siddha, Sowa Rigpa, and Homeopathy---and have their own boards and consensus. Though AYUSH is followed by a large population of Indians, there is limited scientific evidence underpinning these practices, and they can come into conflict with evidence-based medicine. In this paper, by medical consensus, we refer to evidence-based medicine and do not consider AYUSH.
% \chiadd{Chatbots would have a varied, distinct set of users, and the ability of the users to control their interactions and evaluate the beliefs with the bot would not be very consistent. Hence, it is important for the chatbot to consider the possible cultural nuances without depending on the user agency. However, if the bot is targeted and the users are known beforehand, their context and level of comprehension can be fed as input to the bot.}

Given the various complexities, Myna’s decision to rely on a dictionary-based approach and retrieval-augmented generation (RAG) rather than fine-tuning an LLM proved to be more effective in addressing the nuances of cultural sensitivity. Fine-tuning a model can limit its adaptability, as it may not keep pace with the evolving language and behaviors within communities. In contrast, the dictionary was constantly updated with new terms, idioms, and cultural practices as they emerge, ensuring that the chatbot remains relevant and responsive to users' changing needs.
This approach also allows for more control over the chatbot’s responses, ensuring that they align with the cultural and contextual information provided by the knowledge base. Fine-tuning a model could potentially lead to overfitting on outdated or overly rigid cultural norms, while a dictionary-based approach offers greater flexibility and responsiveness. This adaptability is critical for addressing the complexity of healthcare in diverse and dynamic cultural environments.

% s discussed in the methods section, 
\chiadd{Thus far, we have discussed how the chatbot design aims to meet the user's cultural context. However, there is a significant role that user agency plays in shaping how women interpret and make sense of information presented by the chatbot, and act on the information. To integrate their perspectives, Myna prioritized a community-centric methodological approach by ensuring that women from the community played a key role throughout the process, from the pre-development phase to iterative testing and development to evaluation (see Figure \ref{fig:methods}). %Understanding user profiles and contexts can lead to more adaptive AI-driven systems, further promoting user agency \cite{}.
In the evaluation phase, initially, women even had the option to provide open-ended user feedback, and also quantitatively rate the chatbot responses on metrics such as---overall rating, satisfied by answer,  helpful answer, language simplicity, response time, friendliness, and helpfulness. 
% While most metrics were rated on a scale of 1 to 5, satisfied by answer and helpful answer were binary (Yes/No), and user feedback was open-ended.
They consistently rated responses highly, making it challenging to draw meaningful insights. Additionally, some women abstained from rating altogether, and the open-ended feedback option was never utilized. By instead involving the women in focus groups and actively engaging them in generating questions, their agency became increasingly evident. In particular, we noticed a learning effect, where the more time women spent asking questions about SRH, the more complex their queries became over time and their ability to understand medical terms increased.
Though this highlights how individual differences may shape interpretation which can pose a challenge for developers, it also speaks to the women's agency in actively shaping chatbot interactions, during development as well as deployment.}
% This approach highlighted the nature of user agency component, while it can introduce bias through individual perspectives, it empowered the community especially women as active participants in shaping the chatbot’s direction. 

% Ensuring that the chatbot prioritizes user safety while being respectful of cultural sensitivities requires a clear hierarchy of responses that balance medical accuracy with cultural nuance.

% Expecting chatbot to do what humans struggle with in this context. 

% Challenges even in making sense of language used within the community
% E.g. how is abortion vs miscarriage and family planning interpreted in local context? 
% prioritize safety while still being culturally sensitive - establish clear hierarchy

% Users struggle to get the words right to ask the bot and bot sometimes does not understand what user is trying to say. (cyst, copper -t)

% The chatbot provides responses to user queries, which is useful for users seeking health information. However, it sometimes falls short in delivering comprehensive answers, leading to user frustration.
% Users may get frustrated if the bot primarily directs them to consult a doctor without providing additional information. Balancing the recommendation to see a doctor with actionable information is crucial to maintaining user engagement and satisfaction.

%% file: appendix.tex
% \newpage
\section{APPENDIX}
In this section, we include the tables for the description of prompts given to the LLM for response generation and translation, categorization of topics and type of question examples, and original Hinglish text of the chat interactions presented in Section 5 followed by more examples of user queries and chatbot responses.

% \newpage
\begin{table*}[ht]
\centering
%\small
\fontsize{7}{8}\selectfont
   \begin{tabular}{>{\raggedright\arraybackslash}p{2cm} p{2.5cm}p{2cm} p{6.5cm}}
%\begin{tabular}{llll}
\toprule
\chiadd{\textbf{Topic}}& \chiadd{\textbf{Type of Question}} & \chiadd{\textbf{Language}} & \chiadd{\textbf{User Query}} \\
\toprule
\chiadd{Contraceptive methods} & \chiadd{Basic Conceptual Inquiry} & \chiadd{Hinglish} & \chiadd{Condom Kya hota hai?}\\
 & &\chiadd{English
(Translated)}& \chiadd{What is a condom?}\\ 
 \hline
 \chiadd{Family planning} & \chiadd{Complex Queries} & \chiadd{Hinglish} & \chiadd{Family planning ke liye Copper-T lagate hain, vah lagane ke liye kitna time lagta hai??}\\
 & &\chiadd{English
(Translated)}& \chiadd{How much time does it take to insert copper-T for family planning?}\\ 
 \hline
\chiadd{Family planning} & \chiadd{Health/Safety concerns and wellness}  & \chiadd{Hinglish} &\chiadd{ Family planning mein diaphragm use karne se UTI ke problem ho sakte hai kay?} \\
 & &\chiadd{English
(Translated)}& \chiadd{Can using a diaphragm for family planning cause UTI problems?}\\
 \hline
\chiadd{Sexual Health} & \chiadd{Advice/Opinion}  & \chiadd{Hinglish} & \chiadd{agar mai drink kar ke sax karti hu to kya ho sakta hai?} \\
 & &\chiadd{English
(Translated)}& \chiadd{What can happen if I have sex after drinking?}\\
 \hline
 \chiadd{Pregnancy} & \chiadd{Advice/Opinion} & \chiadd{Hinglish} & \chiadd{Konsi aisi bimari hai jis mai hame pregnancy avoid karna chahiye?} \\
 & &\chiadd{English
(Translated)}& \chiadd{What are the diseases during which we should avoid pregnancy?}\\
 \hline
%\chiadd{Pregnancy}& \chiadd{Misconception} & \chiadd{Hinglish} & \chiadd{Papita khane se baccha girta hai kya?} \\
 %& &\chiadd{English
%(Translated)}& \chiadd{Does eating papaya cause a miscarriage?}\\
 %\hline
%\chiadd{Sterilization} & \chiadd{Sexual functionality}& \chiadd{Hinglish} & \chiadd{Mahila nasbandhi ke bad ky sex ki ichha khatm ho jati he ?}\\
 %& &\chiadd{English
%(Translated)}& \chiadd{Does a woman's desire for sex end after sterilization?}\\
 %\hline
%\chiadd{Contraceptive methods} & \chiadd{Menstrual changes} & \chiadd{Hinglish} & \chiadd{Kya Ipill lene ke baad periods jaldi aate hai?}\\
 %& &\chiadd{English
%(Translated)}& \chiadd{Do periods come early after taking an i-pill?}\\
 %\hline
\chiadd{Reproductive anatomy} & \chiadd{Informational} & \chiadd{Hinglish} & \chiadd{Sperm ko kaun control karta hai?} \\
 & &\chiadd{English
(Translated)}& \chiadd{Who controls sperm?}\\
 \hline
%\chiadd{Family planning} & \chiadd{Gender Roles} & \chiadd{Hinglish} & \chiadd{Family planning ki responsibility gents kyu nahi hoti hai?}\\
% & &\chiadd{English
%(Translated)}& \chiadd{Why is family planning not considered the responsibility of men?}\\
 %\hline
\chiadd{Menstrual health} & \chiadd{Health/Safety concerns and wellness} & \chiadd{Hinglish} & \chiadd{Jab period main hote hai to kya chiz nhi khana chahiye} \\
 & &\chiadd{English
(Translated)}& \chiadd{What should not be eaten during periods?}\\
 \hline
\chiadd{Sterilization} & \chiadd{Health/Safety concerns and wellness} & \chiadd{Hinglish} & \chiadd{Mahila nasbandi ke tanke sukhne ke liye Kya Karen?}\\
 & &\chiadd{English
(Translated)}& \chiadd{What should be done to help the stitches from female sterilization heal properly?}\\
 \hline
 \chiadd{Abortion} & \chiadd{Misconception} & \chiadd{Hinglish} & \chiadd{Adrak ka juice peene se kya abortion hota hai?} \\
 & &\chiadd{English
(Translated)}& \chiadd{Does drinking ginger juice cause abortion?}\\
 \hline
 \chiadd{Fertility support} & \chiadd{Informational:Basic Conceptual Inquiry} & \chiadd{Hinglish} & \chiadd{IVF india me bhi hota hai kya?} \\
 & &\chiadd{English
(Translated)}& \chiadd{Is IVF also available in India?}\\
 \hline
 \chiadd{Marriage \& Relationships} & \chiadd{Advice/Opinion} & \chiadd{Hinglish} & \chiadd{Shadi ke bad Hasband ke satha nhi raha hai to Kya  divorce lena sahi hai?}  \\
 & &\chiadd{English
(Translated)}& \chiadd{If I don't stay with my husband after marriage, is it right to get a divorce?}\\
 \hline
 \chiadd{Miscarriage} & \chiadd{Advice/Opinion} & \chiadd{Hinglish} & \chiadd{Pregnancy rukne ke bad bar bar Miscarriage hone ka kya karan ho sakti hai?}  \\
 & &\chiadd{English
(Translated)}& \chiadd{What could be the reasons for repeated miscarriages after conceiving?}\\
 \hline
\chiadd{Family planning} & \chiadd{Norms and Ethics}  & \chiadd{Hinglish} & \chiadd{Religion m operation karna mana hai to kya kare} \\
 & &\chiadd{English
(Translated)}& \chiadd{What to do if religion prohibits surgery?}\\
 \hline
\chiadd{Family planning} & \chiadd{Health Access}  & \chiadd{Hinglish} & \chiadd{Paisa Na Ho to Kya family planning ho sakti hai?} \\
 & &\chiadd{English
(Translated)}& \chiadd{Can family planning be done if there is no money?}\\
 \hline
%\chiadd{Family planning} & \chiadd{Decision Making} & \chiadd{Hinglish} & \chiadd{family planning ke liye kiski Raza Mandi lena zaroori hai ya FIR is kiska Raji hona jaruri hai?} \\
% & &\chiadd{English
%(Translated)}& \chiadd{Whose consent is necessary for family planning, or whose approval is essential for it?}\\
 %\hline
\chiadd{Follow-up} & \chiadd{Follow-up}  & \chiadd{Hinglish} & \chiadd{main doctors se puchna bhul gai thi} \\
 & &\chiadd{English
(Translated)}& \chiadd{I forgot to ask the doctors}\\
\bottomrule
  \end{tabular}
 \caption{\textbf{Categorization of SRH Topics and Type of Questions Covered by Users.} Categorization of topics and the type of questions with examples of how they were grouped.}
     % \vspace{10pt}
  \label{tab:examples}
  \Description{Examples of how the user questions were grouped for topic and type of question are given in both original Hinglish and English text.}
  \end{table*}

\begin{table*}[ht]
\fontsize{7}{8}\selectfont
  \centering
   \begin{tabular}{>{\raggedright\arraybackslash}p{2cm} p{12cm}}
    \toprule
    \textbf{Objective using LLM (GPT-4)} & \textbf{\chiadd{Simplified description of } the prompt given} \\
 \midrule
    Interpret and translate the user query to English & \chirm{You are an expert interpreter and a translator who assists rural Indian women by translating their Hinglish messages related to gynecological health issues into standard English. You should be sensitive to the privacy and cultural context of the users, capable of interpreting common mistakes and conversational and vernacular language, and focused on providing clear, accurate information to improve healthcare communication and outcomes. Please help me interpret and translate the user query considering the things discussed before.

Please return the output in JSON format with the key "translated text".}
\chiadd{A prompt designed to translate colloquial messages into clear English, ensuring respect for cultural nuances and user privacy while preserving the intent of the original message.}
   \\
    \midrule
    Gets the answer from the knowledge base
 & \chirm{You are a seasoned female gynecologist and obstetrician from india, boasting over 25 years of expertise. Ensuring the explanation and sentence formation are easy to understand by a 10-year-old. You understand the importance of cultural sensitivity and will offer you guidance with warmth, empathy, and compassion.

You're guiding rural Indian women, many of whom may lack in-depth biological knowledge and might be apprehensive.

Your core objective is to provide empathetic and dear counsel Prioritize medical accuracy and swiftly correct any misunderstandings cultural nuances and recalling past interactions.

\textbf{Steps :} 
Engage in a step-by-step conversation, mirroring the approach a doctor would take. Identify potential causes for any mentioned ailments. When you need more information, systematically gather patient information, focusing on one follow-up question based on a specific aspect at a time. Give precise guidance to the questions when you understand the user's problem clearly, referring to the information by
our doctors in triple quotes below

"""
\$(vector search results)
"""
When applicable, suggest immediate home-based remedies and precautions.

\textbf{Important rules :}

Avoid recommending tests outright. Ensure your responses. are capped at 150 words, and easily comprehensible. Use easy-to-understand and commonly used words don't use complex medical terms 
Emphasize the importance of consulting a doctor for further diagnosis and treatment. If you perceive reluctance or need to wrap up conversation, direct them to ``Please consult a doctor with Telehealth"

Please return response in JSON format with the key medical answer"}
\chiadd{A prompt crafted to simulate expert guidance in a conversational tone, providing accurate and empathetic answers tailored to user needs. It emphasizes clarity, simplicity, and actionable advice, encouraging professional consultation when appropriate. Responses are concise and accessible.}
  \\
 \midrule
    Fixes the grammar
 & \chirm{Your job is to simplify text and fix grammar without changing the language or script. If the text mixes English with another language or uses a non-traditional script, just simplify and correct grammar. Keep the original meaning, Provide the simplified version in JSON format with 'updated text' as the key.

Simplifying also means replacing words which are heavy in the language by the user.

For example of Hinglish language:

Original message:

Yadi aapka pati ek bachche ke liye tayyar nahi hai, to pitrut ke baare mein aapki bhavnaon, apekshaon aur bhay donon like baare mein ek khull aur imandar baatchit kama mahatvapurn hal

New message:

Agar aapka pati bachche ke liye raazi nahl hal, to bachche aur pita banne ke vishay par aap dono ke vichar, ummeed, aur dar ke baare mein khul kar aur sachchi baat karni zaroori hai.}
\chiadd{A prompt to refine grammar and ensure readability while maintaining the original tone and meaning. It focuses on simplifying language to enhance understanding for the target audience.}
  \\

  \bottomrule
  \end{tabular}
%    \vspace{10pt}
  \caption{\textbf{Prompts given to the LLM.} Response generation and translation prompts given to the LLM.}
    \label{tab:prompt}   
    \Description{Table shows the prompts given to the LLM for response generation and translation. It has two columns with objectives using LLM and a simplified description of the prompt given.}
\end{table*}

\begin{table*}[ht]
 \centering
    \fontsize{7}{8}\selectfont
  \begin{tabular}{>{\raggedright\arraybackslash}p{1.5cm}p{2cm} p{2.5cm} p{8cm}}
    \toprule
 \textbf{Topic}  & \textbf{Language} & \textbf{User Query} & \textbf{Chatbot Response} \\
     \midrule
    \chiadd{Misconceptions and Gender Norms} & Hinglish & Family planning main muje sirf ladka chahiye to uske liye khuch upay hai kya?
  &
   Bacche ka janm chunne ka koi pakka tarika nahi hai. Parivaar niyojan ka mukhya uddeshya garbhavastha ko niyantrit karna hota hai, na ki bacche ke janm ko chunna. Kuch log koshish karte hain ki timing aur tarah-tarah ke tareeke apna kar ladka ya ladki hone ka anumaan lagaye, lekin yeh vidhiyan bilkul bhi pakki nahi hoti hain aur inka koi vaigyanik aadhar nahi hota. Agar aap parivaar niyojan ke vikalpon ke baare mein soch rahe hain, toh aapko un vikalpon par vichar karna chahiye jo aapke aur aapke parivaar ke liye sahi ho. Agar aap aur jaankari chahte hain ya kisi vishesh salah ki zaroorat hai, toh kripya Myna's Telehealth ke saath doctor se salah lein.\\
   \midrule
    \chiadd{Cultural taboo} & Hinglish & Family planning me sex word ko family ke samne kyu Nhi bol na chaya? 
 &
  Har parivaar aur samaj mein baatcheet ke tareeke alag hote hain. Kuch parivaaron mein 'sex' shabd ka istemal khule aam nahi kiya jata kyunki yeh unke liye asahaj ho sakta hai. Yeh samajik ya sanskritik maanyataon ke karan ho sakta hai. Jab family planning ki baat aati hai, toh kai baar log is vishay ko aur adhik samajhdaar tareeke se, jaise 'sambandh banana', 'parivaar niyojan' ya 'bachche ki planning' ke roop mein discuss karte hain. Yeh zaroori hai ki jab aap apne parivaar ke saath is vishay par charcha karte hain, toh aap unki bhavnaon ka samman karein aur aise shabdon ka chayan karein jo sabke liye suvidhajanak ho. Isse baatcheet ko aage badhane mein madad milti hai aur sabhi ko asahaj mehsoos nahi hota.  \\
  \midrule
  \chiadd{Dietary recommendations}& Hinglish & Parivar niyojan nasbandi ke ghav bharane ke liye kya khana chahie
  &
    Nasbandi ke ghav ko bharne ke liye poshtik aahar ka sevan karna bahut zaroori hai. Yahaan kuch aahar sambandhi sujhav diye ja rahe hain jo ghav bharne mein madadgar ho sakte hain:

1. Protein Yukt Aahar: Protein ghav bharne ke liye mahatvapurn hota hai. Daal, chana, soybean, paneer, dahi, ande, chicken, aur machli jaise protein se bharpoor aahar khayein.

2. Vitamin C aur Zinc: Ye dono poshak tatva ghav bharne mein sahayak hote hain. Santre, nimbu, kiwi, tamatar, bell peppers, akhrot, aur beej jaise aahar mein inka sevan badhayein.

3. Harit Sabjiyan: Palak, methi, sarson, aur anya harit sabjiyan vitamins aur minerals se bharpoor hoti hain jo sharir ki marammat mein madad karte hain.

4. Phal: Taaza phal jaise seb, kela, papita, aur berries antioxidants se bharpoor hote hain jo sharir ko swasth rakhte hain.

5. Paryapt Paani: Paryapt matra mein paani piyein kyunki hydration ghav bharne ke liye zaroori hai.

6. Whole Grains: Gehun, jau, aur oats jaise whole grains fiber se bharpoor hote hain jo pachan tantra ko durust rakhte hain.

In aahar ke alava, sharab, caffeine, aur atyadhik chini yukt aahar se bachna chahiye kyunki ye ghav bharne ki prakriya ko dhima kar sakte hain. Kya aapko aur koi sujhav ya jaankari chahiye?\\
\midrule
\chiadd{Religion}& Hinglish & Religion m operation karna mana hai to kya kare &
    Agar aapka dharm ya vishwas operation ya surgical procedures ke khilaf hai, to aapko kuch non-surgical birth control methods ke baare mein sochna chahiye. Kuch non-surgical methods hain:
1. Condoms: Yeh ek prabhavi tarika hai jo pregnancy aur sexually transmitted infections (STIs) se bachav karta hai.
2. Oral contraceptive pills: Mahilao ke liye daily leni padti hain.
3. Vaginal rings: Yeh mahine bhar ke liye insert ki jaati hain.
4. Contraceptive patches: Yeh skin par chipkane wale patches hote hain jo weekly badalne padte hain.
5. Natural family planning methods: Jaise ki ovulation tracking aur withdrawal method.
In sabhi options mein se koi bhi aapke liye sahi ho sakta hai. Lekin, inka istemal karne se pehle, yeh samajhna zaroori hai ki kuch methods jaise ki natural family planning, kam effective ho sakte hain. 
Aapko apne dharmik leader ya samuday ke margdarshan ke saath-saath ek doctor se bhi salah leni chahiye. Aap ``Myna's Telehealth'' se bhi sampark kar sakti hain jahan aapko doctor se salah mil sakti hai jo aapke dharmik vishwas ko samajhte hue aapko sahi margdarshan de sakte hain.\\
   
   \bottomrule
  \end{tabular}
%    \vspace{10pt}
  \caption{\textbf{Taking Social and Cultural Norms into Account (Hinglish).} Chatbot response to questions on cultural beliefs, taboo, food recommendations and role of religion. Find the English text in Table \ref{tab:Baby Sex}}
    \label{tab:stigma}
 \Description{Chatbot response to user questions on cultural beliefs, taboos,
food recommendations, and the role of religion is presented in the table with original Hinglish text. The English text is referenced to Table \ref{tab:Baby Sex}}
  
\end{table*}

\begin{table*}[ht]
 \centering
    \fontsize{7}{8}\selectfont
  \begin{tabular}{>{\raggedright\arraybackslash}p{1.5cm}p{2cm} p{2.5cm} p{8cm}}
    \toprule
    \textbf{Topic} &  \textbf{Language} & \textbf{User Query} & \textbf{Chatbot Response} \\
     
    \midrule
   \chiadd{Age for sex}& Hinglish &  Proper age kya hai first time sex karne ka? &
  Sex ke liye `proper age' ka concept legal aur emotional maturity par depend karta hai. Har desh mein, sex karne ki legal age (jise `age of consent' bhi kaha jata hai) alag hoti hai, aur yeh aam taur par 16 se 18 saal ke beech hoti hai. Ye legal umar yeh sunishchit karni ke liye hoti hai ki vyakti emotionally aur physically mature ho chuke hain apne decisions samajhne aur unke parinaam ko sambhalne ke liye. Lekin, sirf legal age ka hona hi kafi nahi hai. Yeh bhi zaroori hai ki vyakti emotionally ready ho, unke paas sahi jankari ho, aur wo apne partner ke saath ek swasth aur samajhdar rishta banane ke liye taiyar ho. Safe sex aur consent (aapsi sahmati) ke mahatva ko samajhna bhi zaroori hai. Isliye, pehli baar sex karne ka `sahi samay' vyakti ke apne vichar, bhavnaon, aur paristhitiyon par nirbhar karta hai. Agar kisi ko is vishay mein aur madad ya jankari ki zarurat ho, to ek vishwasniya doctor ya healthcare provider se baat karna uchit hoga.  \\
  \midrule
  \chiadd{Family planning}& Hinglish & 15 saal ke ladki family planning kar sakti hai? &
  Haan, 15 saal ki ladki family planning ke baare mein soch sakti hai, lekin is umar mein, zyadatar focus education aur personal development par hota hai. Agar family planning ki zarurat ho, toh sabse pehle non-invasive methods jaise ki barrier methods (condom ka istemal) ya oral contraceptives (goliyan) ke baare mein sochna chahiye. Ye dono tareeke surakshit aur reversible hote hain, matlab aap jab chahein toh inhe band kar sakti hain. Lekin, kisi bhi tarah ka decision lene se pehle, ek visheshagya ya Myna's Telehealth ke doctor se salah zaroor lein. Unka margdarshan aapko sahi disha mein le jayega. \\
  \midrule
   \chiadd{Divorce}&Hinglish & Shadi ke bad Hasband ke satha nhi raha hai to Kya  divorce lena sahi hai? 

  &
   Shaadi ke baad agar aap apne pati ke saath nahi reh paayi hain aur aapko lagta hai ki aapke beech samasyaayein hain, toh pehle yeh samajhna zaroori hai ki samasya kya hai. Har rishta alag hota hai aur kabhi-kabhi samajhauta ya samadhan dhoondhna mushkil ho sakta hai. Lekin, yeh bhi zaroori hai ki aap dono ek dusre se khule mann se baat karein aur samasyaon ka hal dhoondhne ki koshish karein. Agar aapko lagta hai ki aapne sab kuch try kar liya hai aur phir bhi samadhan nahi mil raha, toh aapko apne dil ki sunni chahiye. Yeh ek bahut bada faisla hai, isliye is par salah karne ke liye samay lein aur zaroorat padne par kisi visheshagya ya counsellor se bhi salah lein. Aapke khushi aur swasthya sabse pehle aate hain.\\
   \midrule
   \chiadd{Gender roles}& Hinglish & Purush nasbandi kyun  nahi karte &
   Purush nasbandi, ya vasectomy, ek sthayi garbhnirodhak prakriya hai jo ki bahut surakshit aur prabhavi hoti hai. Lekin, kai baar purush is prakriya ko nahi karwate hain kuch karanon ki wajah se:

1. Jankari ka abhav: Kai purushon ko is prakriya ke baare mein sahi jankari nahi hoti hai.
2. Galatfahmi: Kuch purush sochte hain ki nasbandi se unki yon himmat ya mardangi prabhavit hogi, jo ki galat hai.
3. Samajik Dabav: Samaj mein kai baar aise vichar hote hain ki nasbandi mahilaon ke liye hi hai.
4. Darr ya Bhay: Operation ya surgery ka bhay bhi ek karan ho sakta hai.

Vasectomy ke baad purushon ki yon himmat mein koi kami nahi aati aur na hi unke purush hormones mein koi parivartan hota hai. Yeh ek surakshit aur kam takleefdeh prakriya hai. Agar aap ya aapke parichit is vishay mein aur jaankari chahte hain, toh kripya sthaniya swasthya kendra ya doctor se sampark karein. Aapko aur kuch poochna hai is vishay mein?\\
   \bottomrule
  \end{tabular}
%    \vspace{10pt}
  \caption{\textbf{Responding to Social Dynamics (Hinglish).} Chatbot responses to questions on legal implications, and gender roles. Find the English text in Table \ref{tab:planning}}
    \label{tab:mar}
     \Description{Chatbot response to user questions on legal implications, and gender roles is presented in the table with original Hinglish text. The English text is referenced to Table \ref{tab:planning}}
  
\end{table*}

\begin{table*}[ht]
\centering
\fontsize{7}{8}\selectfont
\begin{tabular}{>{\raggedright\arraybackslash}p{2cm} p{2.5cm} p{9.5cm}}
\toprule
\textbf{Language} & \textbf{User Query} & \textbf{Chatbot Response} \\
\midrule
\multirow{2}{*}{English (Translated)}  & If I'm not getting pregnant, what could be the main reason? & I understand that you might be worried about your problem. There can be several reasons for not getting pregnant. Have you noticed any changes in your menstrual cycle recently? Has your weight increased suddenly or do you have any other health issues like thyroid problems or PCOS? Also, have you ever considered family planning methods? This information will help me better understand your issue. \\
\cmidrule{2-3}
& I don’t know anything, please tell me what could be the reasons. & I understand that you don’t know much about your health, but you’re worried. There could be several reasons for your issue:
\begin{itemize}
    \item \textbf{Hormonal Changes}: Sometimes stress or weight gain/loss can cause hormonal imbalances, leading to irregular periods.
    \item \textbf{Anemia}: Low hemoglobin (iron deficiency) can also cause irregular periods.
    \item \textbf{Thyroid Issues}: Abnormal thyroid gland activity can affect your menstrual cycle.
    \item \textbf{Polycystic Ovary Syndrome (PCOS)}: This condition, where cysts form on the ovaries, can lead to irregular periods.
\end{itemize}
Are you experiencing any other symptoms, such as abdominal pain or mood swings? And have you been involved in any activities that might be causing you stress? This information will help me better understand your problem. \\
\midrule
\multirow{2}{*}{Hinglish} & Agar bacha nahi rukh raha hbai to uska main karan kya ho sakta hai? & Aapki samasya ke liye main samajh sakti hoon ki aapko chinta ho rahi hai. Bacha na rukhne ke kai karan ho sakte hain. Kya aapko pichhle kuch samay se periods mein koi badlav mehsoos ho raha hai? Kya aapka weight sudden mein badh gaya hai ya fir koi aur health issue hai jaise ki thyroid ya PCOS? Aur kya aapne kabhi family planning methods ke baare mein socha hai? Ye jaankari mujhe aapki samasya ko samajhne mein madad karegi. \\
\cmidrule{2-3}
& Mujhe kuch nahi malum muje kya huva hai please aap bataye ki kya karan ho sakte hai & Mujhe samajh hai ki aapko apne swasthya ke baare mein kuch pata nahi hai, lekin aapko chinta toh hai. Aapki samasya ke kuch karan ho sakte hain:
\begin{itemize}
    \item Hormonal badlav: Kabhi-kabhi stress ya wajan badhna/loss ke karan hormones mein asantulan ho sakta hai, jisse periods regular nahi rehte.
    \item Khoon ki kami: Kam hemoglobin (blood mein iron ki kami) bhi periods ko asamanya bana sakta hai.
    \item Thyroid samasya: Thyroid gland ki asamanya gatividhi bhi periods ko prabhavit kar sakti hai.
    \item Polycystic Ovary Syndrome (PCOS): Ismein ovary mein cysts bante hain, jo periods ko asamanya kar dete hain.
\end{itemize}
Aapko kya koi aur lakshan mehsoos ho rahe hain, jaise ki pet mein dard ya chidchidapan? Aur kya aapne koi aisi gatividhi ki hai jisse aapko stress ho raha ho? Ye jaankari mujhe aapki samasya ko samajhne mein madad karegi. \\
\bottomrule
\end{tabular}
\caption{\textbf{Pregnancy query with Follow-up.} Responses to queries regarding reasons for not getting pregnant and follow-up queries and responses for the same.}
\label{tab:reasons}
\Description{Chatbot response to user question regarding reasons for not getting pregnant and follow-up query
and response for the same is presented in a table with both original Hinglish and English text.}
\end{table*}

\begin{table*}[ht]
\centering
\fontsize{7}{8}\selectfont
\begin{tabular}{>{\raggedright\arraybackslash}p{1.5cm} p{2cm} p{2.5cm} p{8cm}}
\toprule
\textbf{Topic}&\textbf{Language} & \textbf{User Query} & \textbf{Chatbot Response} \\
\midrule
 Employment& English(Translated) & How important is one's job for family planning? &
My work is very important in family planning. I provide women and their partners with accurate
information and guidance about their family planning options. This helps them decide when and
how many children they want. I would tell them about different types of contraceptives, like the pill,
IUD, or condoms, and also explain their advantages, disadvantages, and contraindications. Apart from
this, I also give them advice regarding pregnancy related problems and their solutions. My aim is that
every person and every family should make informed decisions regarding their health and future. In
this way, I can contribute to improving the health and well-being of the society.\\
\cmidrule{2-4}
  &Hinglish &Apni job family planning ke liye kitni important hai?&Mera kaam parivar niyojan mein bahut mahatvapurna hai. Main mahilaon aur unke sathiyon ko unke parivar niyojan ke vikalpon ke baare mein sahi jankari aur margadarshan pradan karta hoon. Ye unhe ye tay karne mein madad karta hai ki woh kab aur kitne bachche chahte hain. Main unhe alag-alag garbh nirodhak ke baare mein batata hoon, jaise ki goliyan, IUD, ya Condom, aur unke fayde, nuksan, aur dushparinam ke baare mein bhi samjhaata hoon. Iske alava, main unhe garbhavastha se sambandhit samasyaon aur unke samadhanon ke baare mein bhi salah deta hoon. Mera uddeshya hai
ki har vyakti aur har parivar apne swasthya aur bhavishya ke nirnay soch-samajh kar le. Is tarah,
main samaj mein swasthya aur kalyan ko behtar dene mein yogdan deta hoon.\\
\bottomrule
\end{tabular}
\caption{\textbf{Understanding Information-seeking behaviors around SRH.} Chatbot response to a question on the relationship between having a job and family planning.}
\label{tab:emp}
\Description{User query and chatbot response on the relationship between job and family planning are given in both original Hinglish and English text.}
  
\end{table*}